\documentclass{article}
\usepackage{amsmath}

\numberwithin{equation}{section}
\input{tcilatex}
\begin{document}

\centerline{\large\bf SLOCC classification for nine families of
four-qubits \footnote{The paper was supported by NSFC(Grants No.
60433050, 60673034 and 10875061)
 }}

\centerline{Dafa Li \footnote{email
address:dli@math.tsinghua.edu.cn}}

\centerline{ Dept of mathematical sciences, Tsinghua University,
Beijing 100084 CHINA}

\centerline{Xiangrong Li }

\centerline{ Department of Mathematics, University of California, 
Irvine, CA 92697-3875, USA}

\centerline{Hongtao Huang}

\centerline{ Electrical Engineering and Computer Science Department} %
\centerline{ University of Michigan, Ann Arbor, MI 48109, USA}

\centerline{Xinxin Li} 
\centerline{ Dept. of computer science, Wayne
State University, Detroit, MI 48202, USA}

Abstract

In Phys. Rev. A 62, 062314 (2000), D\"{u}r, Vidal and Cirac indicated that
there are infinitely many SLOCC classes for four qubits. Verstraete,
Dehaene, and Verschelde in Phys. Rev. A 65, 052112 (2002) proposed nine
families of states corresponding to nine different ways of entangling four
qubits. In Phys. Rev. A 75, 022318 (2007), Lamata et al. reported that there
are eight true SLOCC entanglement classes of four qubits up to permutations
of the qubits. In this paper, we investigate SLOCC classification of the
nine families proposed by Verstraete, Dehaene and Verschelde, and
distinguish 49 true SLOCC entanglement classes from them.



\section{Introduction}

Recently, many authors have exploited SLOCC (stochastic local operations and
classical communication ) classification. D\"{u}r et al. showed that for
pure states of three qubits there are four different degenerate SLOCC
entanglement classes and two different true entanglement classes \cite{Dur},
and for pure states of four qubits, there are infinitely many SLOCC classes.
Their proof is briefly described here. A general state of $n$ qubits depends
on $2\times 2^{n}-2$ real parameters\ in \cite{Dur}. In \cite{Dur}, the
invertible local operator $\alpha $ was fixed to that $\det (\alpha )=1$.
Hence, the local operator $\alpha $ depends on six real parameters. Thus,
the set of SLOCC\ classes of $n$ qubits depends on at least $2\times
2^{n}-2-6n$ real parameters. Then, it was asserted \cite{Dur} there exist
finite SLOCC classes for three qubits while infinite SLOCC classes for $%
n\geq 4$ qubits. The argument is not complete due to the following two
reasons. (i). By the definition, for any invertible operator $\alpha $, $%
\det (\alpha )\neq 0$. The operator $\alpha $ with $\det (\alpha )\neq 0$\
still depends on eight real parameters because we can image to remove a
hyperplane $\det (\alpha )=0$ with six real parameters from a space with
eight real parameters. That is, the set of SLOCC\ classes of $n$ qubits
depends on at least $2\times 2^{n}-2-8n$ real parameters. By the argument in 
\cite{Dur}, this lower bound allows for a finite number of SLOCC classes for 
$n=4$. (ii). In Sec. 7.1 of this paper, we show that two different true
SLOCC entanglement\ classes constitute a continuous Family $L_{a_{4}}$ in 
\cite{Moor2}. The state $L_{a_{4}}$ with $a=0$ and the state $L_{a_{4}}$
with $a=1$ represent these two classes. Clearly, the latter class including
the state $L_{a_{4}}$ with $a\neq 0$ can be labeled by a continuous
parameter $a\neq 0$. It means that finite parameters may allow for finite
SLOCC classes. In other words, it cannot be asserted that there exist
infinite SLOCC classes for four qubits even though the set of SLOCC\ classes
of four qubits depends on at least six real parameters.

Verstraete et al. \cite{Moor2} proposed that for four qubits, there exist
nine families of states corresponding to nine different ways of
entanglement. They gave a representative state for each family and claimed
that by determinant-one SLOCC operations, a pure state of four qubits can be
transformed into one of the nine families up to permutations of the qubits.

In \cite{Lamata1}\cite{Lamata2}, the authors used the partition to
investigate SLOCC\ classification of three qubits and four qubits. The idea
for the partition was originally used to analyze the separability of $n$
qubits and multipartite pure states in \cite{LDF}. In \cite{Lamata2}, the
authors reported that there are 16 true SLOCC entanglement classes of four
qubits, where permutation is explicitly included in the counting. Up to
permutations of the qubits, there are eight true SLOCC entanglement classes
of four qubits. We can show that this classification is not complete. For
example, for Span $\{O_{k}\Psi ,O_{k}\Psi \}$ in \cite{Lamata2}, the
canonical states are

$|0000\rangle +|1100\rangle +a|0011\rangle +b|1111\rangle $ and

$|0000\rangle +|1100\rangle $ $+a|0001\rangle +a|0010\rangle +b|1101\rangle
+b|1110\rangle $,

\noindent where $a\neq b$ \cite{Lamata2}. It was pointed out in \cite{LDF07a}
that for the former canonical state, $a=-b$ and $a\neq -b$ represent two
different true SLOCC entanglement classes, while for the latter canonical
state, $ab=0$ and $ab\neq 0$ represent another two different true SLOCC
entanglement classes. It says that the partition approach cannot classify
the two subfamilies represented by the above canonical states under SLOCC.
Using the method in this paper, we can classify each Span $\{.....\}$ in 
\cite{Lamata2}. In total, the eight Spans $\{...\}$ in \cite{Lamata2}\
include more than 16 true SLOCC entanglement classes.

Miyake proposed the onionlike classification of SLOCC\ orbits \cite{Miyake03}%
. The simple criteria for the complete SLOCC\ classification for three
qubits were given in \cite{LDF-PLA}. In \cite{LDF07}, we proposed the SLOCC
invariants and semi-invariants for four qubits. Using the invariants and
semi-invariants, it can be determined if two states belong to different
SLOCC entanglement classes. In \cite{LDF07a}, in terms of invariants and
semi-invariants\ we distinguished 28 distinct true entanglement classes of
four qubits, where permutations of the qubits are allowed. That
classification is not complete. The invariants and semi-invariants only
require simple arithmetic operations. In this paper, we will investigate
SLOCC classification of each family in \cite{Moor2} by means of invariants
and semi-invariants. We want to know how many SLOCC entanglement classes
there are for each family by the definition in \cite{Dur}. For example, in
this paper we show that Family $L_{a_{4}}$ only has two SLOCC entanglement
classes: $L_{a_{4}}$ with $a=0$ and $L_{a_{4}}$ with $a\neq 0$, which both
are true entanglement classes. The class $L_{a_{4}}$ with $a\neq 0$\
includes a continuous parameter of $SL$ (determinant-one SLOCC) classes. We
also show that Family $L_{a_{2}0_{3\oplus 1}}$ only has two SLOCC
entanglement classes: $L_{a_{2}0_{3\oplus 1}}$ with $a=0$ and $%
L_{a_{2}0_{3\oplus 1}}$ with $a\neq 0$, and the latter is a true
entanglement class. As well, we demonstrate that the class $%
L_{a_{2}0_{3\oplus 1}}$ with $a\neq 0$\ includes a continuous parameter of $%
SL$ classes, and can be labeled by a continuous parameter $a\neq 0$.

In this paper, we distinguish at least 49 true SLOCC entanglement classes
from Verstraete et al.'s nine families. For example, Family $G_{abcd}$, $13$%
; Family $L_{abc_{2}}$, $19$; Family $L_{a_{2}b_{2}}$, $4$; Family $%
L_{ab_{3}}$, $8$; Family $L_{a_{4}}$, $2$; Family $L_{a_{2}0_{3\oplus 1}}$, $%
1$; Family $L_{0_{5\oplus 3}}$, $1$; Family $L_{0_{7\oplus 1}}$, $1$; Family 
$L_{0_{3+\bar{1}}0_{3+\bar{1}}}$, $0$. We give the complete SLOCC\
classifications for families $L_{a_{4}}$, $L_{a_{2}0_{3\oplus 1}}$, $%
L_{0_{5\oplus 3}}$, $L_{0_{7\oplus 1}}$, and $L_{0_{3+\bar{1}}0_{3+\bar{1}}}$%
. But we cannot guarantee that the SLOCC classifications for other families
are complete.

In Sec. 3 of this paper, we exploit the classification for Family $G_{abcd}$%
. In Sec. 4 of this paper, we discuss the classification for Family $%
L_{abc_{2}}$. In Sec. 5 and Sec. 6 of this paper, we study the
classification for families $L_{a_{2}b_{2}}$ and $L_{ab_{3}}$, respectively.
For the classifications of other families, see Sec. 7.

For the readability, we list the definitions of invariants and
semi-invariants in Sec. 2 of this paper.

\section{SLOCC invariants and semi-invariants}

The states of a four-qubit system can be generally expressed as 
\begin{equation}
|\psi \rangle =\sum_{i=0}^{15}a_{i}|i\rangle .
\end{equation}

By definition in \cite{Dur}, two states $|\psi \rangle $ and $|\psi ^{\prime
}\rangle $ are equivalent under SLOCC if and only if there exist invertible
local operators $\alpha ,\beta $, $\gamma $ and $\delta $\ such that

\begin{equation}
|\psi ^{\prime }\rangle =\alpha \otimes \beta \otimes \gamma \otimes \delta
|\psi \rangle ,  \label{SLOCC-1}
\end{equation}

\noindent where the local operators $\alpha ,\beta ,$ $\gamma $ and $\delta $
can be expressed as $2\times 2$ invertible matrices

\begin{equation*}
\alpha =\left( 
\begin{array}{cc}
\alpha _{1} & \alpha _{2} \\ 
\alpha _{3} & \alpha _{4}%
\end{array}%
\right) ,~\beta =\left( 
\begin{array}{cc}
\beta _{1} & \beta _{2} \\ 
\beta _{3} & \beta _{4}%
\end{array}%
\right) ,~\gamma =\left( 
\begin{tabular}{cc}
$\ \gamma _{1}$ & $\ \gamma _{2}$ \\ 
$\ \gamma _{3}$ & $\ \gamma _{4}$%
\end{tabular}%
\right) ,\delta =\left( 
\begin{tabular}{cc}
$\ \delta _{1}$ & $\ \delta _{2}$ \\ 
$\ \delta _{3}$ & $\ \delta _{4}$%
\end{tabular}%
\right) .
\end{equation*}

\subsection{SLOCC invariant}

Let $|\psi ^{\prime }\rangle =\sum_{i=0}^{15}a_{i}^{\prime }|i\rangle $ in
Eq. (\ref{SLOCC-1}). If $|\psi ^{\prime }\rangle $ is SLOCC\ equivalent to $%
|\psi \rangle $, then the following equation was derived by induction in 
\cite{LDF07}.

\begin{equation}
\mathcal{I}(\psi ^{\prime })=\mathcal{I}(\psi )\det (\alpha )\det (\beta
)\det (\gamma )\det (\delta ),  \label{IV0}
\end{equation}

where

\begin{equation}
\mathcal{I}(\psi
)=(a_{0}a_{15}-a_{1}a_{14})-(a_{2}a_{13}-a_{3}a_{12})-(a_{4}a_{11}-a_{5}a_{10})+(a_{6}a_{9}-a_{7}a_{8})
\label{IV1}
\end{equation}%
and 
\begin{equation}
\mathcal{I}(\psi ^{\prime })=(a_{2}^{\prime }a_{13}^{\prime }-a_{3}^{\prime
}a_{12}^{\prime })+(a_{4}^{\prime }a_{11}^{\prime }-a_{5}^{\prime
}a_{10}^{\prime })-(a_{0}^{\prime }a_{15}^{\prime }-a_{1}^{\prime
}a_{14}^{\prime })-(a_{6}^{\prime }a_{9}^{\prime }-a_{7}^{\prime
}a_{8}^{\prime }).  \label{IV2}
\end{equation}

Notice that $\mathcal{I}(\psi )$ does not vary under $SL$-operations, i.e.,
determinant-one SLOCC\ operations, or vanish under\textbf{\ }%
non-determinant-one SLOCC operations. It is easy to see that if $|\psi
^{\prime }\rangle $ and $|\psi \rangle $ are equivalent under SLOCC, then
either $\mathcal{I}(\psi ^{\prime })=\mathcal{I}(\psi )=0$ or $\mathcal{I}%
(\psi ^{\prime })\mathcal{I}(\psi )\neq 0$. Otherwise, the two states belong
to different SLOCC entanglement classes. Especially, if $\mathcal{I}(\psi
^{\prime })\neq \mathcal{I}(\psi )$ , then $|\psi \rangle $ and $|\psi
^{\prime }\rangle $ are inequivalent under $SL$-operations. Eq. (\ref{IV0})
implies that each SLOCC entanglement class has infinite $SL$-classes. This
is also true for $n$ qubits.

By solving matrix equation in Eq. (\ref{SLOCC-1}), we obtain the amplitudes $%
a_{i}^{\prime }$ of $|\psi ^{\prime }\rangle $. By substituting $%
a_{i}^{\prime }$ into the above $\mathcal{I}(\psi ^{\prime })$ in Eq. (\ref%
{IV2}), we obtain the values of $\mathcal{I}(\psi ^{\prime })$ in the Tables
of this paper.

\subsection{Semi-invariants $D_{1}$, $D_{2}$ and $D_{3}$}

In \cite{LDF-PLA}\cite{LDF07a}, we defined $D_{i}(\psi )$ for the state $%
|\psi \rangle $ as follows.

\begin{eqnarray}
D_{1}(\psi
)=(a_{1}a_{4}-a_{0}a_{5})(a_{11}a_{14}-a_{10}a_{15})-(a_{3}a_{6}-a_{2}a_{7})(a_{9}a_{12}-a_{8}a_{13}),
\label{D-1} \\
D_{2}(\psi
)=(a_{4}a_{7}-a_{5}a_{6})(a_{8}a_{11}-a_{9}a_{10})-(a_{0}a_{3}-a_{1}a_{2})(a_{12}a_{15}-a_{13}a_{14}),
\label{D-2} \\
D_{3}(\psi
)=(a_{3}a_{5}-a_{1}a_{7})(a_{10}a_{12}-a_{8}a_{14})-(a_{2}a_{4}-a_{0}a_{6})(a_{11}a_{13}-a_{9}a_{15}).
\label{D-3}
\end{eqnarray}

Let $|\psi \rangle $ be the representative states in Tables in this paper.\
By solving matrix equation in Eq. (\ref{SLOCC-1}), we obtain the amplitudes $%
a_{i}^{\prime }$ of $|\psi ^{\prime }\rangle $. By substituting $%
a_{i}^{\prime }$ into $D_{1}$, $D_{2}$, and $D_{3}$ in Eqs. (\ref{D-1},\ref%
{D-2},\ref{D-3}), we obtain the values of $D_{1}$, $D_{2}$, and $D_{3}$ in
these Tables. If $D_{i}=0$ for some class in these Tables, then it implies
that $D_{i}=0$ for for all the states of that class. If $D_{i}$ is $\Delta $%
\ for some class in these Tables, then it means that $D_{i}=0$ for some
states of that class while $D_{i}\neq 0$ for other states of that class. For
example, for the class A1.2 in Table I, $D_{1}$ is $0$, $D_{2}$ is $\Delta $%
,\ and $D_{3}=0$. It says that for some state of the class A1.2 in Table I, $%
D_{2}\neq 0$ while $D_{1}=D_{3}=0$ for each state of the class A1.2.

\subsection{Semi-invariants $F_{i}$}

$F_{1}(\psi
)=(a_{0}a_{7}-a_{2}a_{5}+a_{1}a_{6}-a_{3}a_{4})^{2}-4(a_{2}a_{4}-a_{0}a_{6})(a_{3}a_{5}-a_{1}a_{7}), 
$

$F_{2}(\psi
)=(a_{8}a_{15}-a_{11}a_{12}+a_{9}a_{14}-a_{10}a_{13})^{2}-4(a_{11}a_{13}-a_{9}a_{15})(a_{10}a_{12}-a_{8}a_{14}), 
$

$F_{3}(\psi
)=(a_{0}a_{11}-a_{2}a_{9}+a_{1}a_{10}-a_{3}a_{8})^{2}-4(a_{2}a_{8}-a_{0}a_{10})(a_{3}a_{9}-a_{1}a_{11}), 
$

$F_{4}(\psi
)=(a_{4}a_{15}-a_{6}a_{13}+a_{5}a_{14}-a_{7}a_{12})^{2}-4(a_{6}a_{12}-a_{4}a_{14})(a_{7}a_{13}-a_{5}a_{15}), 
$

$F_{5}(\psi
)=(a_{0}a_{13}-a_{4}a_{9}+a_{1}a_{12}-a_{5}a_{8})^{2}-4(a_{4}a_{8}-a_{0}a_{12})(a_{5}a_{9}-a_{1}a_{13}), 
$

$F_{6}(\psi
)=(a_{2}a_{15}-a_{6}a_{11}+a_{3}a_{14}-a_{7}a_{10})^{2}-4(a_{6}a_{10}-a_{2}a_{14})(a_{7}a_{11}-a_{3}a_{15}), 
$

$F_{7}(\psi
)=(a_{0}a_{14}-a_{4}a_{10}+a_{2}a_{12}-a_{6}a_{8})^{2}-4(a_{4}a_{8}-a_{0}a_{12})(a_{6}a_{10}-a_{2}a_{14}), 
$

$F_{8}(\psi
)=(a_{1}a_{15}-a_{5}a_{11}+a_{3}a_{13}-a_{7}a_{9})^{2}-4(a_{5}a_{9}-a_{1}a_{13})(a_{7}a_{11}-a_{3}a_{15}), 
$

$F_{9}(\psi
)=(a_{0}a_{15}-a_{2}a_{13}+a_{1}a_{14}-a_{3}a_{12})^{2}-4(a_{0}a_{14}-a_{2}a_{12})(a_{1}a_{15}-a_{3}a_{13}), 
$

$F_{10}(\psi
)=(a_{4}a_{11}-a_{7}a_{8}+a_{5}a_{10}-a_{6}a_{9})^{2}-4(a_{7}a_{9}-a_{5}a_{11})(a_{6}a_{8}-a_{4}a_{10})) 
$. \ \ 

Let $|\psi \rangle $ be the representative states in Tables in this paper.\
By solving matrix equation in Eq. (\ref{SLOCC-1}), we obtain the amplitudes $%
a_{i}^{\prime }$ of $|\psi ^{\prime }\rangle $. By substituting $%
a_{i}^{\prime }$ into the above $F_{i}$, we can obtain the values of $F_{i}$%
. Thus, we can derive the properties of $F_{i}$ for a SLOCC class.

Denotation. Let $P=\det^{2}(\beta )\det^{2}(\delta )\det^{2}(\gamma )$, $%
Q=\det^{2}(\alpha )\det^{2}(\gamma )\det^{2}(\delta )$, $R=\det^{2}(\alpha
)\det^{2}(\beta )\det^{2}(\delta )$, $S=\det^{2}(\alpha )\det^{2}(\gamma
)\det^{2}(\beta )$, $T=\det (\alpha )\det (\beta )\det (\gamma )\det (\delta
)$.

\section{Family $G_{abcd}$}

\ The representative state of this family is $G_{abcd}=$\ $\frac{a+d}{2}%
(|0000\rangle +|1111\rangle )+\frac{a-d}{2}(|0011\rangle +|1100\rangle )+%
\frac{b+c}{2}(|0101\rangle +|1010\rangle )+\frac{b-c}{2}(|0110\rangle
+|1001\rangle )$. $G_{abcd}$ becomes a product state of two $EPR$ pairs for
the following cases:

$a=b=c=d$; $x=y=z=0$ and $u\neq 0$; $x=y=z=-u$; $x=y=-z=-u$, where $x$, $y$, 
$z$, $u$ are distinct and $x$, $y$, $z$, $u\in \{a,b,c,d\}$.

When either $b=c=0$ and $a=\pm d\neq 0$ or $a=d=0$ and $b=\pm c\neq 0$, the
states obtained from $G_{abcd}$ belong to the class $|GHZ\rangle $.

The state $G_{abcd}$ satisfies the following equations. 
\begin{eqnarray}
&&\mathcal{I}=(a^{2}+b^{2}+c^{2}+d^{2})/2,  \notag \\
&&D_{1}=(ac+bd)(ab+cd)/4,  \notag \\
&&D_{2}=(a^{2}+b^{2}-c^{2}-d^{2})(-a^{2}+b^{2}-c^{2}+d^{2})/16,  \notag \\
&&D_{3}=-(-ac+bd)(ab-cd)/4,  \notag \\
&&F_{i}=(b^{2}-c^{2})(a^{2}-d^{2})/4,i=1\text{ to }8,  \notag \\
&&F_{9}=a^{2}d^{2},F_{10}=b^{2}c^{2}.  \label{Gabcd-eq}
\end{eqnarray}

We consider the following four subfamilies and list the true SLOCC
entanglement classes of each subfamily in Tables I (1), I (2.1), I (2.2) and
I (2.3).

\subsection{Subfamily $G_{abcd}$ with $x=y=0$ and $zu\neq 0$, where
different $x$, $y$, $u$, $v\in \{a$, $b$, $c$, $d\}$}

(1). $G_{abcd}$ with $a=d=0$ and $bc\neq 0$ is equivalent to $G_{abcd}$ with 
$b=c=0$ and $ad\neq 0$ under SLOCC $\sigma _{x}\otimes I\otimes I\otimes
\sigma _{x}$, where $I$ is the identity matrix.

(2). $G_{abcd}$ with $c=d=0$ and $ab\neq 0$ is equivalent to $G_{abcd}$ with 
$a=b=0$ and $cd\neq 0$ under SLOCC $I\otimes \sigma _{y}\otimes \sigma
_{y}\otimes I$.

(3). $G_{abcd}$ with $b=d=0$ and $ac\neq 0$ is equivalent to $G_{abcd}$ with 
$a=c=0$ and $bd\neq 0$ under SLOCC $I\otimes \sigma _{x}\otimes I\otimes
\sigma _{x}$.

(4). $G_{abcd}$ with $b=d=0$ and $ac\neq 0$ is equivalent to $G_{abcd}$ with 
$c=d=0$ and $ab\neq 0$. This is because $a(|0000\rangle +|1111\rangle
+|0011\rangle +|1100\rangle )+c(|0101\rangle +|1010\rangle -|0110\rangle
-|1001\rangle )$

$=\alpha \otimes \beta \otimes \gamma \otimes \delta (-a(|0000\rangle
+|1111\rangle +|0011\rangle +|1100\rangle )+c(|0101\rangle +|1010\rangle
+|0110\rangle +|1001\rangle ))$, where $\alpha =\gamma =diag\{i$, $1\}$, $%
\beta =diag\{-i$, $1\}=-\delta $.

Hence, we only need to consider the subfamily $G_{abcd}$ with $b=c=0$ and $%
ad\neq 0$.\ In this subfamily $G_{abcd}$ with $b=c=0$ and $ad\neq 0$, there
are three true SLOCC entanglement classes, denoted as A1.1 ( i.e., $%
|GHZ\rangle $), A1.2, and A1.3.

For the class A1.1, it includes states $G_{abcd}$ with $b=c=0$ and $a=\pm
d\neq 0$, which are equivalent to $|GHZ\rangle $.

For the class A1.2, it includes states $G_{abcd}$ with $b=c=0$ and $%
a^{2}+d^{2}=0$. A1.2 is a true SLOCC entanglement class. We can argue this
as follows. It is straightforward to verify $G_{abcd}(b=c=0,d=\pm ai)=\alpha
\otimes \beta \otimes \gamma \otimes \delta $ $G_{abcd}(b=c=0,a=1,d=\pm i)$,
where $\alpha =\delta =diag(\sqrt{a},1)$ and $\beta =\gamma =diag(1,\sqrt{a}%
) $. Also, $G_{abcd}(b=c=0,a=1,d=-i)=\alpha \otimes \beta \otimes \gamma
\otimes \delta G_{abcd}(b=c=0,a=1,d=i)$, where $\alpha =diag(i,1)$, $\beta
=diag(1,-i)$, $\gamma =diag(-1,1)$, and $\delta =I$.

Verstraete indicated that Family $G_{abcd}$ includes the state $|\phi
_{4}\rangle =$ $(|0000\rangle +|0011\rangle +|1100\rangle -|1111\rangle )/2$%
. Now, we can exactly say that the state $|\phi _{4}\rangle $ is in the
class A1.2 whose representative is the state $G_{abcd}(b=c=0,a=1,d=i)$. This
is because the representative state is equivalent to $|\phi _{4}\rangle $
under SLOCC $\alpha \otimes \beta \otimes \gamma \otimes \delta $, where $%
\alpha =diag(i,1)$, $\beta =I$, $\gamma =diag(1-i,1)$ and $\delta
=diag(1,-(1+i))$.

For the class A1.3, it includes states $G_{abcd}$ with $b=c=0$ and $%
a^{2}+d^{2}\neq 0$. We cannot classify A1.3 further.

The three classes A1.1, A1.2 and A1.3 are different by the values of $%
\mathcal{I}$, $D_{1}$, $D_{2}$, and $D_{3}$ in Table I (1).

For each state of this subfamily, the following equations hold.

$\mathcal{I}=(a^{2}+d^{2})/2\ast T$, $D_{1}=0$, $D_{2}=(a^{2}-d^{2})/16\ast
(...)$, $D_{3}=0$,

$F_{1}=\allowbreak a^{2}d^{2}\alpha _{1}^{2}\alpha _{2}^{2}\ast P$, $%
F_{2}=a^{2}d^{2}\alpha _{3}^{2}\alpha _{4}^{2}\ast P$, $F_{3}=a^{2}d^{2}%
\beta _{1}^{2}\beta _{2}^{2}\ast Q$, $F_{4}=\allowbreak a^{2}d^{2}\beta
_{3}^{2}\beta _{4}^{2}\ast Q$,

$F_{5}=a^{2}d^{2}\gamma _{2}^{2}\gamma _{1}^{2}\ast R$, $F_{6}=\allowbreak
a^{2}d^{2}\gamma _{4}^{2}\gamma _{3}^{2}\ast R$, $F_{7}=\allowbreak
a^{2}d^{2}\delta _{1}^{2}\delta _{2}^{2}\ast S$, $F_{8}=a^{2}d^{2}\delta
_{3}^{2}\delta _{4}^{2}\ast S$,

$F_{9}=\allowbreak a^{2}d^{2}\left( \alpha _{1}\beta _{1}\alpha _{4}\beta
_{4}-\beta _{3}\alpha _{3}\beta _{2}\alpha _{2}\right) ^{2}\det^{2}(\delta
)\det^{2}(\gamma )$,

$F_{10}=a^{2}d^{2}\left( -\alpha _{1}\beta _{3}\alpha _{4}\beta _{2}+\beta
_{1}\alpha _{3}\beta _{4}\alpha _{2}\right) ^{2}\allowbreak \det^{2}(\delta
)\det^{2}(\gamma )$.

From the above equations, we have the following property.

Property 1.1.

For each state of this subfamily $G_{abcd}$ with $b=c=0$ and $ad\neq 0$, if $%
{\small F_{1}F_{2}=0}\wedge {\small F}_{3}{\small F}_{4}={\small 0}$ then $%
{\small F}_{9}{\small =0}\wedge {\small F}_{10}{\small \neq 0}$ or ${\small F%
}_{9}{\small \neq 0}\wedge {\small F}_{10}{\small =0}.$

By computing, we obtain the Table I (1). For the first row of Table I (1),
each state of the class $|GHZ\rangle $ must satisfy $\mathcal{I}\neq 0$, $%
D_{1}=D_{2}=D_{3}=0$. $\Delta $ in this paper may be zero or not.

Table I (1). The true SLOCC entanglement classes in the subfamily $G_{abcd}$
with $b=c=0$ and $ad\neq 0$

\begin{tabular}{|l|l|l|l|l|l|}
\hline
classes & criteria for classification & $\mathcal{I}$ & $D_{1}$ & $D_{2}$ & $%
D_{3}$ \\ \hline
A1.1 (i.e., $|GHZ\rangle $) & $a=\pm d$ & $\neq 0$ & $0$ & $0$ & $0$ \\ 
\hline
A1.2 (i.e., $|\phi _{4}\rangle $) & $a\neq \pm d$, and $a^{2}+d^{2}=0$ & $=0$
& $0$ & $\Delta $ & $0$ \\ \hline
A1.3, * & $a\neq \pm d$, and $a^{2}+d^{2}\neq 0$ & $\neq 0$ & $0$ & $\Delta $
& $0$ \\ \hline
\end{tabular}

Note that \textquotedblleft *\textquotedblright\ means that we cannot show
that it has only one true SLOCC entanglement class.

\subsection{Subfamily $G_{abcd}$ with $x=\pm y\neq 0$ and $u=\pm v\neq 0$,
where different $x$, $y$, $u$, $v\in \{a$, $b$, $c$, $d\}$}

\subsubsection{Subsubfamily $G_{abcd}$ with $a=\pm d\neq 0$ and $b=\pm c\neq
0$}

(1). The state $G_{abcd}$ with $a=-d$ and $b=c$ is equivalent to $G_{abcd}$
with $a=d$ and $b=-c$ under SLOCC $I\otimes \sigma _{x}\otimes I\otimes $\ $%
\sigma _{x}$.

(2). The state $G_{abcd}$ with $a=-d$ and $b=-c$ is equivalent to $G_{abcd}$
with $a=d$ and $b=c$ under SLOCC $I\otimes I\otimes \sigma _{x}\otimes $\ $%
\sigma _{x}$.

Therefore we only need to consider the subsubfamily $G_{abcd}$ with $a=d$
and $b=\pm c$ but $a\neq \pm b$ as follows.

This subsubfamily has four true SLOCC entanglement classes denoted as A2.1,
A2.2, A3.1 and A3.2.

For the class A2.1, it includes states $G_{abcd}$ with $a=d$, $b=c,$ $a\neq
\pm b$, and $a^{2}+b^{2}=0$. That is, A2.1 includes $a(|0000\rangle
+|1111\rangle )\pm ai(|0101\rangle +|1010\rangle )$. A2.1 is a true SLOCC
entanglement class. We can argue this as follows. It is straightforward to
verify $a(|0000\rangle +|1111\rangle )\pm ai(|0101\rangle +|1010\rangle
)=\alpha \otimes \beta \otimes \gamma \otimes \delta $ $(|0000\rangle
+|1111\rangle \pm i(|0101\rangle +|1010\rangle ))$, where $\alpha =\delta
=diag(\sqrt{a},1)$ and $\beta =\gamma =diag(1,\sqrt{a})$. We can also show
that two states $(|0000\rangle +|1111\rangle \pm i(|0101\rangle
+|1010\rangle ))$\ are equivalent under SLOCC. This is because $%
(|0000\rangle +|1111\rangle -i(|0101\rangle +|1010\rangle ))\ =\alpha
\otimes \beta \otimes \gamma \otimes \delta $ $(|0000\rangle +|1111\rangle
+i(|0101\rangle +|1010\rangle ))$, where $\gamma =\delta =I$ and $\alpha
=\beta =\sigma _{z}$.

For the class A2.2, it includes states $G_{abcd}$ with $a=d$, $b=c$, $a\neq
\pm b$, and $a^{2}+b^{2}\neq 0$. We cannot classify A2.2 further.

For the class A3.1, it includes states $G_{abcd}$ with $a=d$, $b=-c$, $a\neq
\pm b$, and $a^{2}+b^{2}=0$. That is, A3.1 includes $a(|0000\rangle
+|1111\rangle )\pm ai(|0110\rangle +|1001\rangle )$. We can show that A3.1
is a true SLOCC entanglement class as follows. It is plain to verify that $%
a(|0000\rangle +|1111\rangle )\pm ai(|0110\rangle +|1001\rangle )=\alpha
\otimes \beta \otimes \gamma \otimes \delta $ $(|0000\rangle +|1111\rangle
\pm i(|0110\rangle +|1001\rangle ))$, where $\alpha =\beta =diag(\sqrt{a},1)$
and $\gamma =\delta =diag(1,\sqrt{a})$. Furthermore. we can demonstrate that 
$(|0000\rangle +|1111\rangle +i(|0110\rangle +|1001\rangle ))=\alpha \otimes
\beta \otimes \gamma \otimes \delta (|0000\rangle +|1111\rangle
-i(|0110\rangle +|1001\rangle ))$, where $\gamma =\delta =I$ and $\alpha
=\beta =\sigma _{z}$.

For the class A3.2, it includes states $G_{abcd}$ with $a=d$, $b=-c$, $a\neq
\pm b$, and $a^{2}+b^{2}\neq 0$. We cannot classify A3.2 further.

The four classes are different by the values of $\mathcal{I}$, $D_{1}$, $%
D_{2}$, and $D_{3}$ in Table I (2.1).

Table I (2.1). The true SLOCC entanglement classes in the subsubfamily $%
G_{abcd}$ with $a=d$ and $b=\pm c$ but $a\neq \pm b$

\begin{tabular}{|l|l|l|l|l|l|}
\hline
classes & criteria for classification & $\mathcal{I}$ & $D_{1}$ & $D_{2}$ & $%
D_{3}$ \\ \hline
A2.1 & $b=c$, $a^{2}+b^{2}=0$ & $=0$ & $\Delta $ & $0$ & $0$ \\ \hline
A2.2, * & $b=c$, $a^{2}+b^{2}\neq 0$ & $\neq 0$ & $\Delta $ & $0$ & $0$ \\ 
\hline
A3.1 & $b=-c$, $a^{2}+b^{2}=0$ & $=0$ & $0$ & $0$ & $\Delta $ \\ \hline
A3.2, * & $b=-c$, $a^{2}+b^{2}\neq 0$ & $\neq 0$ & $0$ & $0$ & $\Delta $ \\ 
\hline
\end{tabular}

Let us show that this subsubfamily is different from the subfamily $G_{abcd}$
with $x=y=0$ and $zu\neq 0$ as follows.

The representative state of this subsubfamily is $G_{abcd}$ with $a=d$ and $%
b=\pm c$. From Eq. (\ref{Gabcd-eq}), this representative state satisfies $%
F_{i}=0$, $i=1$ to $8$, $F_{9}=a^{4}$, and $F_{10}=b^{4}$. This violates
property 1.1. Therefore this subsubfamily is different from the subfamily $%
G_{abcd}$ with $x=y=0$ and $zu\neq 0$.

By calculating, each state of classes A2.1 and A2.2 satisfies the following
equations.

$\mathcal{I}=(a^{2}+b^{2})$, $D_{1}=-ab\ast (...)$, $D_{2}=D_{3}=0$,

$F_{1}=(a^{2}-b^{2})\allowbreak \alpha _{1}^{2}\alpha _{2}^{2}\ast P$, $%
F_{2}=(a^{2}-b^{2})\alpha _{3}^{2}\alpha _{4}^{2}\ast P$,

$F_{3}=(a^{2}-b^{2})\beta _{1}^{2}\beta _{2}^{2}\ast Q$, $F_{4}=\allowbreak
(a^{2}-b^{2})\beta _{3}^{2}\beta _{4}^{2}\ast Q$,

$F_{5}=(a^{2}-b^{2})\gamma _{2}^{2}\gamma _{1}^{2}\ast R$, $%
F_{6}=(a^{2}-b^{2})\allowbreak \gamma _{4}^{2}\gamma _{3}^{2}\ast R$,

$F_{7}=(a^{2}-b^{2})\allowbreak \delta _{1}^{2}\delta _{2}^{2}\ast S$, $%
F_{8}=(a^{2}-b^{2})\delta _{3}^{2}\delta _{4}^{2}\ast S$,

We omit $F_{9}$ and $F_{10}$ because they are too complicated. From the
above $F_{i}$, we can derive the following property 1.2.

Property 1.2.

(1). For each state of the subsubfamily $G_{abcd}$ with $a=d$ and $b=c$ but $%
a\neq \pm b$,

\ \ (i). if $\mathcal{I}=0$, then $F_{9}=F_{10}$,

\ \ (ii). if ${\small F_{1}F_{2}=0}\wedge {\small F}_{3}{\small F}_{4}=%
{\small 0}$, then ${\small F}_{9}{\small \neq 0}\wedge {\small F}_{10}%
{\small \neq 0}$,

\ \ (iii). if $\mathcal{I}\neq 0$ and ${\small F_{1}F_{2}=0}\wedge {\small F}%
_{3}{\small F}_{4}={\small 0}$, then ${\small F}_{9}{\small \neq F}_{10}$.

(2). For each state of the subsubfamily $G_{abcd}$ with $a=d$ and $b=c$ but $%
a\neq \pm b$, if ${\small F_{1}F_{2}=0\wedge F}_{3}={\small F}_{4}={\small 0}
$ or ${\small F}_{3}{\small F}_{4}={\small 0\wedge F_{1}=F_{2}=0}$, then $%
D_{1}\neq 0$.

\subsubsection{Subsubfamily $G_{abcd}$ with $a=\pm b\neq 0$, and $c=\pm
d\neq 0$}

(1). When $a=b\wedge $ $c=d$ or $a=-b\wedge $ $c=-d$, and $a^{2}+c^{2}=0$,
the states are in A2.1.

(2). When $a=b\wedge $ $c=d$ or $a=-b\wedge $ $c=-d$, and $a^{2}+c^{2}\neq 0$%
, the states are in A2.2.

(3). When $a=b\wedge c=-d$ or $a=-b\wedge c=d$, and $a^{2}+c^{2}=0$, the
states are in A3.1.

(4). When $a=b\wedge c=-d$ or $a=-b\wedge c=d$, and $a^{2}+c^{2}\neq 0$, the
states are in A3.2.

\subsubsection{Subsubfamily $G_{abcd}$ with $a=\pm c\neq 0$ and $b=\pm d\neq
0$}

The states $G_{abcd}$ with $a=\pm c$ and $b=\pm d$ are equivalent to the
states $G_{abcd}$ with $a=\pm b$, and $c=\pm d$ in the above subsection
3.2.2 under SLOCC. For example, the state $G_{abcd}$ with $a=c$ and $b=d$,
denoted as $G_{abcd}(a=c$ and $b=d)$, is equivalent to the state $G_{abcd}$
with $a=b$, and $c=d$, denoted as $G_{abcd}(a=b$ and $c=d)$. This fact can
be verified as follows. Let $\alpha =\delta =diag\{i$, $1\}$, $\beta =\gamma
=diag\{-i$, $1\}$. Then, it is easy to see that $G_{abcd}(a=b$ and $%
c=d)=\alpha \otimes \beta \otimes \gamma \otimes \delta G_{abcd}(a=c$ and $%
b=d)$.

\subsection{Subfamily $G_{abcd}$ with either $a=\pm d\neq 0$ and $b\neq \pm
c $, or $b=\pm c\neq 0$ and $a\neq \pm d$}

Note that

(1). The state $G_{abcd}$ with $b=c$ and $a\neq \pm d$\ is equivalent to
state $G_{abcd}$\ with $a=d$ and $b\neq \pm c$ under $\sigma _{x}\otimes
I\otimes \sigma _{x}\otimes I$.

(2). The state $G_{abcd}$ with $b=-c$ and $a\neq \pm d$\ is equivalent to
the state $G_{abcd}$ with $a=d$ and $b\neq \pm c$ under $\sigma _{x}\otimes
I\otimes I\otimes \sigma _{x}$.

(3). Furthermore, the state $G_{abcd}$\ with $a=-d$ and $b\neq \pm c$ is
equivalent to the state $G_{abcd}$\ with $a=d$ and $b\neq \pm c$ under $%
I\otimes I\otimes \sigma _{x}\otimes \sigma _{x}$.

Therefore, we only need to consider the subfamily $G_{abcd}$ with $a=d$ and $%
b\neq \pm c$ as follows.

This subfamily has four true SLOCC entanglement classes denoted as A4.1,
A4.2, A4.3 and A4.4.

For the class A4.1, it includes states $G_{abcd}$ with $a=d$, either $a=\pm
b $ or $\pm c$, and $2a^{2}+b^{2}+c^{2}=0$.

For the class A4.2, it includes states $G_{abcd}$ with $a=d$, either $a=\pm
b $ or $\pm c$, and $2a^{2}+b^{2}+c^{2}\neq 0$.

For the class A4.3, it includes states $G_{abcd}$ with $a=d$, $a\neq \pm b$, 
$a\neq $ $\pm c$, and $2a^{2}+b^{2}+c^{2}=0$.

For the class A4.4, it includes states $G_{abcd}$ with $a=d$, $a\neq \pm b$, 
$a\neq $ $\pm c$, and $2a^{2}+b^{2}+c^{2}\neq 0$.

We cannot classify A4.1, A4.2, A4.3 or A4.4 further.

Table I (2.2). The true SLOCC entanglement classes in the subfamily $%
G_{abcd} $ with $a=d$ and $b\neq \pm c$

\begin{tabular}{|l|l|l|l|l|l|}
\hline
classes & criteria for classification & $\mathcal{I}$ & $D_{1}$ & $D_{2}$ & $%
D_{3}$ \\ \hline
A4.1, * & either $a=\pm b$ or $\pm c$, and $2a^{2}+b^{2}+c^{2}=0$ & $=0$ & $%
\Delta $ & $\Delta $ & $\Delta $ \\ \hline
A4.2, * & either $a=\pm b$ or $\pm c$, and $2a^{2}+b^{2}+c^{2}\neq 0$ & $%
\neq 0$ & $\Delta $ & $\Delta $ & $\Delta $ \\ \hline
A4.3, * & $a\neq \pm b$, $a\neq \pm c$, and $2a^{2}+b^{2}+c^{2}=0$ & $=0$ & $%
\Delta $ & $\Delta $ & $\Delta $ \\ \hline
A4.4, * & $a\neq \pm b$, $a\neq \pm c$, and $2a^{2}+b^{2}+c^{2}\neq 0$ & $%
\neq 0$ & $\Delta $ & $\Delta $ & $\Delta $ \\ \hline
\end{tabular}

We demonstrate the following properties of this subfamily.

When $a=d$, $G_{abcd}$ becomes $G_{abcd}(a=d)=a(|0000\rangle +|1111\rangle )+%
\frac{b+c}{2}(|0101\rangle +|1010\rangle )+\frac{b-c}{2}(|0110\rangle
+|1001\rangle )$. Any state connected with $G_{abcd}(a=d)$ by SLOCC
satisfies the following: \ 

$\mathcal{I}=1/2(2a^{2}+b^{2}+c^{2})T$,

\begin{eqnarray}
F_{1} &=&\allowbreak (a^{2}-b^{2})\allowbreak (a^{2}-c^{2})^{2}\alpha
_{1}^{2}\alpha _{2}^{2}P,F_{2}=\allowbreak (a^{2}-b^{2})\allowbreak
(a^{2}-c^{2})^{2}\alpha _{3}^{2}\alpha _{4}^{2}P,  \notag \\
F_{3} &=&\allowbreak (a^{2}-b^{2})\allowbreak (a^{2}-c^{2})^{2}\beta
_{1}^{2}\beta _{2}^{2}Q,F_{4}=\allowbreak (a^{2}-b^{2})\allowbreak
(a^{2}-c^{2})^{2}\allowbreak \beta _{3}^{2}\beta _{4}^{2}Q,  \notag \\
F_{5} &=&\allowbreak (a^{2}-b^{2})\allowbreak (a^{2}-c^{2})^{2}\gamma
_{2}^{2}\gamma _{1}^{2}R,F_{6}=\allowbreak (a^{2}-b^{2})\allowbreak
(a^{2}-c^{2})^{2}\allowbreak \gamma _{4}^{2}\gamma _{3}^{2}R,  \notag \\
F_{7} &=&\allowbreak (a^{2}-b^{2})\allowbreak (a^{2}-c^{2})^{2}\allowbreak
\delta _{1}^{2}\delta _{2}^{2}S,F_{8}=\allowbreak (a^{2}-b^{2})\allowbreak
(a^{2}-c^{2})^{2}\delta _{3}^{2}\delta _{4}^{2}S.  \label{F-values}
\end{eqnarray}

We omit the complicated expressions of $D_{1}$, $D_{2}$, $D_{3}$, $F_{9}$
and $F_{10}$.

From Eq. (\ref{F-values}), we have the following property 1.3.

Property 1.3.

(1). From Eq. (\ref{F-values}),\ each state of the subfamily $G_{abcd}$ with 
$a=d$, $b\neq \pm c$, and either $a=\pm b$ or $\pm c$ satisfies $F_{i}=0$, $%
i=1$ to $8$, and when $b^{2}+c^{2}\neq 0$, $\ \left\vert F_{9}\right\vert
+\left\vert F_{10}\right\vert \neq 0$.

(2). From Eq. (\ref{F-values}),\ for each state of the subfamily $G_{abcd}$
with $a=d$, $b\neq \pm c$, $a\neq \pm b$ and $a\neq \pm c$, if $%
F_{1}=F_{2}=F_{3}=F_{4}=0$, then one can obtain (i) $F_{9}\neq 0$ and $%
F_{10}\neq 0$ whenever $abcd\neq 0$, and (ii) $F_{9}\neq $ $F_{10}$ whenever 
$a^{4}\neq b^{2}c^{2}$.

From Eq. (\ref{F-values}), $F_{i}\neq 0$, $i=1$, ..., $8$ for some states of
the subfamily $G_{abcd}$ with $a=d$, $b\neq \pm c$. Therefore, classes A4.1,
A4.2, A4.3 and A4.4 are different from each other. Also see Table I (2.2).

Let us argue that this subfamily is different from the subfamily $G_{abcd}$
with $x=y=0$ and $zu\neq 0$ and the subfamily $G_{abcd}$ with $a=\pm d$ and $%
b=\pm c$ as follows.

The representative state of this subfamily is $G_{abcd}$ with $a=d$ and $%
b\neq \pm c$. From Eq. (\ref{Gabcd-eq}), this representative state satisfies 
$F_{i}=0$, $i=1$ to $8$, $F_{9}=a^{4}$, $F_{10}=b^{2}c^{2}$, $%
D_{1}=a^{2}(b+c)^{2}/4\neq 0$, $D_{2}=(b^{2}-c^{2})^{2}/16\neq 0$, $%
D_{3}=-a^{2}(b-c)^{2}/4\neq 0$. These $F_{i}$ violate property 1.1. Hence,
this subfamily is different from the subfamily $G_{abcd}$ with $x=y=0$ and $%
zu\neq 0$. These $D_{i}$ also violate the criteria for $D_{i}$ in Table I
(2.1), therefore this subfamily is different from the subfamily $G_{abcd}$
with $a=\pm d$ and $b=\pm c$.

\subsection{Subfamily $G_{abcd}$ with $x\neq \pm y$, or $x\neq \pm y$\ but
only one $r=s$,\ where $x$, $y\in \{a$, $b$, $c$, $d\}$, $r\in \{\pm a$, $%
\pm d\}$, and $s\in \{\pm b$, $\pm c\}$}

Class A4.5 includes states $G_{abcd}$ with $a^{2}+b^{2}+c^{2}+d^{2}=0$. The
class A4.6 includes states $G_{abcd}$ with $a^{2}+b^{2}+c^{2}+d^{2}\neq 0$.
We cannot classify A4.5 or A4.6 further.

Table I (2.3). The true SLOCC entanglement classes in the subfamily $%
G_{abcd} $ with $x\neq \pm y$

\begin{tabular}{|l|l|l|l|l|l|}
\hline
classes & criteria for classification & $\mathcal{I}$ & $D_{1}$ & $D_{2}$ & $%
D_{3}$ \\ \hline
A4.5, * & $a^{2}+b^{2}+c^{2}+d^{2}=0$ & $=0$ & $\Delta $ & $\Delta $ & $%
\Delta $ \\ \hline
A4.6, * & $a^{2}+b^{2}+c^{2}+d^{2}\neq 0$ & $\neq 0$ & $\Delta $ & $\Delta $
& $\Delta $ \\ \hline
\end{tabular}

For this subfamily, $\mathcal{I}$ $=\frac{1}{2}(a^{2}+b^{2}+c^{2}+d^{2})T$.
The two classes are different because of their different values of $\mathcal{%
I}$. See Table I (2.3).

\section{Family $L_{abc_{2}}$}

The representative state of this family is $L_{abc_{2}}=$\ $\frac{a+b}{2}%
(|0000\rangle +|1111\rangle )+\frac{a-b}{2}(|0011\rangle +|1100\rangle )+c(\
|0101\rangle +|1010\rangle )+|0110\rangle $. When $a=b=c=0$, this becomes a
full separable state. We divide Family $L_{abc_{2}}$\ into three
subfamilies. They are the subfamily $L_{abc_{2}}$ with $c=0$, the subfamily $%
L_{abc_{2}}$ with $abc\neq 0$, and the subfamily $L_{abc_{2}}$ with $c\neq 0$%
, $ab=0$. We list the true SLOCC entanglement classes of each subfamily in
Tables II (1), II (2), II (3), and II (4), and demonstrate that these
classes of each subfamily are distinct true SLOCC entanglement classes in
Appendix A.

\subsection{The classification for the subfamily $L_{abc_{2}}$ with $c=0$}

Here the state $L_{abc_{2}}$ with $c=0$\ represents this subfamily. Let $%
L_{abc_{2}}(c=0)=\frac{a+b}{2}(|0000\rangle +|1111\rangle )+\frac{a-b}{2}%
(|0011\rangle +|1100\rangle )+|0110\rangle $. The subfamily $%
L_{abc_{2}}(c=0) $ includes the following five distinct true SLOCC
entanglement classes. The five classes are denoted as B1.1, B1.2, B1.3,
B1.4, and B1.5. See Table II (1).

For the class B1.1, it includes states $L_{abc_{2}}$ with $c=0$ and $a=b\neq
0$. That is, B1.1 includes $a(|0000\rangle +|1111\rangle )+|0110\rangle $.
B1.1 is a true SLOCC entanglement class because $a(|0000\rangle
+|1111\rangle )+|0110\rangle =\alpha \otimes \beta \otimes \gamma \otimes
\delta (|0000\rangle +|1111\rangle +|0110\rangle )$, where $\gamma =\delta
=I $, $\alpha =diag(1,a)$, and $\beta =diag(a,1)$.

For the class B1.2, it includes states $L_{abc_{2}}$ with $c=0$ and $%
a=-b\neq 0$. That is, B1.2 includes $a(|0011\rangle +|1100\rangle
)+|0110\rangle $. B1.2 is a true SLOCC entanglement class because $%
a(|0011\rangle +|1100\rangle )+|0110\rangle =\alpha \otimes \beta \otimes
\gamma \otimes \delta (|0011\rangle +|1100\rangle +|0110\rangle )$, where $%
\gamma =\delta =I$, $\alpha =diag(1,a)$, and $\beta =diag(a,1)$.

For the class B1.3, it includes states $L_{abc_{2}}$ with $c=0$ and $a\neq
\pm b$ and $a^{2}+b^{2}=0$. That is, B1.3 includes $\frac{a(1\pm i)}{2}%
(|0000\rangle +|1111\rangle )+\frac{a(1\mp i)}{2}(|0011\rangle +|1100\rangle
)+|0110\rangle $. We can prove that B1.3 is a true SLOCC entanglement class
as follows. It is easy to verify that $\frac{a(1\pm i)}{2}(|0000\rangle
+|1111\rangle )+\frac{a(1\mp i)}{2}(|0011\rangle +|1100\rangle
)+|0110\rangle =\alpha \otimes \beta \otimes \gamma \otimes \delta (\frac{%
(1\pm i)}{2}(|0000\rangle +|1111\rangle )+\frac{(1\mp i)}{2}(|0011\rangle
+|1100\rangle )+|0110\rangle )$, where $\alpha =diag(1,a)$, $\beta
=diag(a,1) $, $\gamma =\delta =I$. Also, we can show that $\frac{(1+i)}{2}%
(|0000\rangle +|1111\rangle )+\frac{(1-i)}{2}(|0011\rangle +|1100\rangle
)+|0110\rangle =\alpha \otimes \beta \otimes \gamma \otimes \delta (\frac{%
(1-i)}{2}(|0000\rangle +|1111\rangle )+\frac{(1+i)}{2}(|0011\rangle
+|1100\rangle )+|0110\rangle )$, where $\alpha =diag(1,i)$, $\beta
=diag(-i,1)$, $\gamma =diag(-1,1)$, and $\delta =I$.

For the class B1.4, it includes states$\ L_{abc_{2}}$ with $c=0$, $a\neq \pm
b$, $ab\neq 0$, and $a^{2}+b^{2}\neq 0$. We cannot classify B1.4 further.

For the class B1.5, it includes states $L_{abc_{2}}$ with $c=0$, $a\neq \pm
b $, and $ab=0$. B1.5 is a true SLOCC entanglement class. The following is
our argument. $\frac{a}{2}(|0000\rangle +|1111\rangle )\pm \frac{a}{2}%
(|0011\rangle +|1100\rangle )+|0110\rangle =\alpha \otimes \beta \otimes
\gamma \otimes \delta (\frac{1}{2}(|0000\rangle +|1111\rangle )\pm \frac{1}{2%
}(|0011\rangle +|1100\rangle )+|0110\rangle )$, where $\alpha =diag(1,a)$, $%
\beta =diag(a,1)$, $\gamma =\delta =I$. We can also show $\frac{1}{2}%
(|0000\rangle +|1111\rangle )+\frac{1}{2}(|0011\rangle +|1100\rangle
)+|0110\rangle =\alpha \otimes \beta \otimes \gamma \otimes \delta (\frac{1}{%
2}(|0000\rangle +|1111\rangle )-\frac{1}{2}(|0011\rangle +|1100\rangle
)+|0110\rangle )$, where $\alpha =\delta =\sigma _{z}$, $\beta =$ $\gamma =I$%
.

Classes B1.1-B1.5 are distinct because of their different values of $%
\mathcal{I}$,$\ D_{1}$, $D_{2}$, and $D_{3}$ in Table II (1).

Table II (1). The true SLOCC entanglement classes in the subfamily $%
L_{abc_{2}}$ with $c=0$

\begin{tabular}{|l|l|l|l|l|l|}
\hline
classes & criteria for classification & $\mathcal{I}$ & $D_{1}$ & $D_{2}$ & $%
D_{3}$ \\ \hline
B1.1 & $a=b\neq 0$ & $\neq 0$ & $0$ & $0$ & $\Delta $ \\ \hline
B1.2 & $a=-b\neq 0$ & $\neq 0$ & $\Delta $ & $0$ & $0$ \\ \hline
B1.3 & $a\neq \pm b$, and $a^{2}+b^{2}=0$ & $=0$ & $\Delta $ & $\Delta $ & $%
\Delta $ \\ \hline
B1.4, * & $a\neq \pm b$, $ab\neq 0$, and $a^{2}+b^{2}\neq 0$ & $\neq 0$ & $%
\Delta $ & $\Delta $ & $\Delta $ \\ \hline
B1.5 & $a\neq \pm b$, and $ab=0$ & $\neq 0$ & $0$ & $\Delta $ & $0$ \\ \hline
\end{tabular}

\subsection{The classification for the subfamily $L_{abc_{2}}$ with $abc\neq
0$}

This subfamily has three inequivalent subsubfamilies under SLOCC. They are
the subsubfamily $L_{abc_{2}}$ with $abc\neq 0$ and $a=b$, the subsubfamily $%
L_{abc_{2}}$ with $abc\neq 0$ and $a=-b$, and the subsubfamily $L_{abc_{2}}$
with $abc\neq 0$, $a\neq \pm b$.

Table II (2). The true SLOCC entanglement classes in the subfamily $%
L_{abc_{2}}$ with $abc\neq 0$

\begin{tabular}{|l|l|l|l|l|l|}
\hline
classes & criteria for classification & $\mathcal{I}$ & $D_{1}$ & $D_{2}$ & $%
D_{3}$ \\ \hline
B2.1 & $a=b$, $a=\pm c$ & $\neq 0$ & $\Delta $ & $0$ & $0$ \\ \hline
B2.2 & $a=b$, $a\neq \pm c$, and $a^{2}+c^{2}=0$ & $=0$ & $\Delta $ & $%
\Delta $ & $\Delta $ \\ \hline
B2.3, * & $a=b$, $a\neq \pm c$, and $a^{2}+c^{2}\neq 0$ & $\neq 0$ & $\Delta 
$ & $\Delta $ & $\Delta $ \\ \hline
B3.1 & $a=-b$, and $a=\pm c$ & $\neq 0$ & $0$ & $0$ & $\Delta $ \\ \hline
B3.2 & $a=-b$, $a\neq \pm c$, and $a^{2}+c^{2}=0$ & $=0$ & $\Delta $ & $%
\Delta $ & $\Delta $ \\ \hline
B3.3, * & $a=-b$, $a\neq \pm c$, and $a^{2}+c^{2}\neq 0$ & $\neq 0$ & $%
\Delta $ & $\Delta $ & $\Delta $ \\ \hline
B4.1, * & $a\neq \pm b$, $a=c$, and $(3a^{2}+b^{2})=0$ & $=0$ & $\Delta $ & $%
\Delta $ & $\Delta $ \\ \hline
B4.2, * & $a\neq \pm b$, $a=c$, and $(3a^{2}+b^{2})\neq 0$ & $\neq 0$ & $%
\Delta $ & $\Delta $ & $\Delta $ \\ \hline
\end{tabular}

\subsubsection{Subsubfamily $L_{abc_{2}}$ with $abc\neq 0$ and $a=b$}

This subsubfamily has the following three distinct true SLOCC entanglement
classes. They are named as B2.1, B2.2, and B2.3. See Table II (2).

For the class B2.1,\ it includes states $L_{abc_{2}}$ with $abc\neq 0$, $a=b$%
, and $a=\pm c$. That is, B2.1 includes $a$ $(|0000\rangle +|1111\rangle
)\pm a$\ $($\ $|0101\rangle +|1010\rangle )+|0110\rangle $. We can show that
B2.1 is a true entanglement class as follows.\ $a$ $(|0000\rangle
+|1111\rangle )\pm a$\ $($\ $|0101\rangle +|1010\rangle )+|0110\rangle
=\alpha \otimes \beta \otimes \gamma \otimes \delta (|0000\rangle
+|1111\rangle \pm \ (|0101\rangle +|1010\rangle )+|0110\rangle )$, where $%
\alpha =diag(1,a)$, $\gamma =diag(a,1)$, $\beta =\delta =I$. Also, $%
|0000\rangle +|1111\rangle +\ |0101\rangle +|1010\rangle +|0110\rangle
=\alpha \otimes \beta \otimes \gamma \otimes \delta (|0000\rangle
+|1111\rangle -\ (|0101\rangle +|1010\rangle )+|0110\rangle )$, where $%
\alpha =\delta =\sigma _{z}$, and $\beta =\gamma =I$.

For the class B2.2, it includes states $L_{abc_{2}}$ with $abc\neq 0$, $a=b$%
, $a\neq \pm c$, and $a^{2}+c^{2}=0$. That is, B2.2 includes $a$ $%
(|0000\rangle +|1111\rangle )\pm ai$\ $($\ $|0101\rangle +|1010\rangle
)+|0110\rangle $. We can also show that B2.2 is a true entanglement class as
follows. $a(|0000\rangle +|1111\rangle )\pm ai$\ $($\ $|0101\rangle
+|1010\rangle )+|0110\rangle =\alpha \otimes \beta \otimes \gamma \otimes
\delta (|0000\rangle +|1111\rangle \pm i\ (|0101\rangle +|1010\rangle
)+|0110\rangle )$, where $\alpha =diag(1,a)$, $\gamma =diag(a,1)$, $\beta
=\delta =I$. Also, $|0000\rangle +|1111\rangle +i$\ $($\ $|0101\rangle
+|1010\rangle )+|0110\rangle =\alpha \otimes \beta \otimes \gamma \otimes
\delta (|0000\rangle +|1111\rangle -i\ (|0101\rangle +|1010\rangle
)+|0110\rangle )$, where $\alpha =\delta =\sigma _{z}$, and $\beta =\gamma
=I $.

For the class B2.3, it includes states $L_{abc_{2}}$ with $abc\neq 0$, $a=b$%
, $a\neq \pm c$, and $a^{2}+c^{2}\neq 0$. We cannot classify B2.3 further.

Classes B2.1, B2.2, and B2.3 are different from each other by the values of $%
\mathcal{I}$,$\ D_{1}$, $D_{2}$, and $D_{3}$ in Table II (2).

\subsubsection{The classification for the subsubfamily $L_{abc_{2}}$ with $%
abc\neq 0$ and $a=-b$}

This subsubfamily has the following three distinct true SLOCC entanglement
classes. They are denoted as B3.1, B3.2, and B3.3. See Table II (2).

For the class B3.1, it includes states $L_{abc_{2}}$ with $abc\neq 0$ and $%
a=-b$ and $a=\pm c$. That is, B3.1 includes $a$ $(|0011\rangle +|1100\rangle
)\pm a$\ $($\ $|0101\rangle +|1010\rangle )+|0110\rangle $. We demonstrate
that B3.1 is a true entanglement class below.

$a$ $(|0011\rangle +|1100\rangle )\pm a$\ $($\ $|0101\rangle +|1010\rangle
)+|0110\rangle =\alpha \otimes \beta \otimes \gamma \otimes \delta
(|0011\rangle +|1100\rangle \pm \ (|0101\rangle +|1010\rangle )+|0110\rangle
)$, where $\alpha =\delta =I$, $\beta =\gamma =diag(a,1)$.

Also, $|0011\rangle +|1100\rangle +\ |0101\rangle +|1010\rangle
+|0110\rangle =\alpha \otimes \beta \otimes \gamma \otimes \delta
(|0011\rangle +|1100\rangle -\ (|0101\rangle +|1010\rangle )+|0110\rangle )$%
, where $\alpha =\gamma =I$, $\beta =\sigma _{z}$, and $\delta =-\sigma _{z}$%
.

For the class B3.2, it includes states $L_{abc_{2}}$ with $abc\neq 0$ and $%
a=-b$ and $a\neq \pm c$, $a^{2}+c^{2}=0$. That is, B3.2 includes$\ a$ $%
(|0011\rangle +|1100\rangle )\pm ai$\ $($\ $|0101\rangle +|1010\rangle
)+|0110\rangle $. We can argue that B3.2 is a true entanglement class as
follows.

$a$ $(|0011\rangle +|1100\rangle )\pm ai$\ $($\ $|0101\rangle +|1010\rangle
)+|0110\rangle =\alpha \otimes \beta \otimes \gamma \otimes \delta
(|0011\rangle +|1100\rangle \pm i\ (|0101\rangle +|1010\rangle
)+|0110\rangle )$, where $\alpha =\delta =I$, $\beta =\gamma =diag(a,1)$.

Also, $|0011\rangle +|1100\rangle +i\ (|0101\rangle +|1010\rangle
)+|0110\rangle =\alpha \otimes \beta \otimes \gamma \otimes \delta
(|0011\rangle +|1100\rangle -i\ (|0101\rangle +|1010\rangle )+|0110\rangle )$%
, where $\alpha =\gamma =I$, $\beta =\sigma _{z}$, and $\delta =-\sigma _{z}$%
.

For the class B3.3, it includes states $L_{abc_{2}}$ with $abc\neq 0$ and $%
a=-b$ and $a\neq \pm c$, $a^{2}+c^{2}\neq 0$. We cannot classify B3.3
further.

Classes B3.1, B3.2 and B3.3 are different from each other by the values of $%
\mathcal{I}$,$\ D_{1}$, $D_{2}$, and $D_{3}$ in Table II (2).

\subsubsection{Subsubfamily $L_{abc_{2}}$ with $abc\neq 0$, $a\neq \pm b$,
but $c=\pm a$ or $c=\pm b$}

Note that the states $L_{abc_{2}}$ with $abc\neq 0$, $a\neq \pm b$, and $%
c=\pm b$\ can be\ obtained by SLOCC $I\otimes \sigma _{z}\otimes \sigma
_{z}\otimes I$\ from the states $L_{abc_{2}}$\ with $abc\neq 0$, $a\neq \pm
b $, and $c=\mp a$, respectively.

Furthermore, the state $L_{abc_{2}}$\ with $abc\neq 0$, $a\neq \pm b$, and $%
c=-a$ is equivalent to the $L_{abc_{2}}$\ with $abc\neq 0$, $a\neq \pm b$,
and $c=a$ under SLOCC. The following is our argument. Let $p=-a$ and $q=-b$.
Then, from the state $\frac{p+q}{2}(|0000\rangle +|1111\rangle )+\frac{p-q}{2%
}(|0011\rangle +|1100\rangle )+p(\ |0101\rangle +|1010\rangle )+|0110\rangle 
$, we can obtain $\frac{a+b}{2}(|0000\rangle +|1111\rangle )+\frac{a-b}{2}%
(|0011\rangle +|1100\rangle )-a(\ |0101\rangle +|1010\rangle )+|0110\rangle $
under SLOCC $(-\sigma _{z})\otimes \sigma _{z}\otimes I\otimes I$.

Therefore, we only need to consider states $L_{abc_{2}}$ with $abc\neq 0$, $%
a\neq \pm b$, $a=c$. This subsubfamily has the following two different true
SLOCC entanglement classes: B4.1 and B4.2. See Table II (2).

For the class B4.1, it includes states $L_{abc_{2}}$ with $abc\neq 0$, $%
a\neq \pm b$, $c=a$ and $(3a^{2}+b^{2})=0$.

For the class B4.2, it includes states $L_{abc_{2}}$ with $abc\neq 0$, $%
a\neq \pm b$, $a=c$, $(3a^{2}+b^{2})\neq 0$.

We cannot classify B4.1 or B4.2 further.

Classes B4.1 and B4.2 are different because they have different values of $%
\mathcal{I}$ in Table II (2).

\subsubsection{Subsubfamily $L_{abc_{2}}$ with $abc\neq 0$, and $x\neq \pm y$%
, where $x$, $y\in \{a$, $b$, $c\}$}

We can distinguish two classes B4.3 and B4.4 from this subsubfamily.

For the class B4.3, it includes states $L_{abc_{2}}$ with $abc\neq 0$ and $%
x\neq \pm y$ and $a^{2}+b^{2}+2c^{2}=0$.

For the class B4.4, it includes states $L_{abc_{2}}$ with $abc\neq 0$ and $%
x\neq \pm y$ and $a^{2}+b^{2}+2c^{2}\neq 0$.

We cannot classify B4.3 or B4.4 further.

Table II (3). The true SLOCC entanglement classes in the subfamily $%
L_{abc_{2}}$ with $abc\neq 0$ and $x\neq \pm y$

\begin{tabular}{|l|l|l|l|l|l|}
\hline
classes & criteria for classification & $\mathcal{I}$ & $D_{1}$ & $D_{2}$ & $%
D_{3}$ \\ \hline
B4.3, * & $x\neq \pm y$, $a^{2}+b^{2}+2c^{2}=0$ & $=0$ & $\Delta $ & $\Delta 
$ & $\Delta $ \\ \hline
B4.4, * & $x\neq \pm y$, $a^{2}+b^{2}+2c^{2}\neq 0$ & $\neq 0$ & $\Delta $ & 
$\Delta $ & $\Delta $ \\ \hline
\end{tabular}

These two classes are different because of their different values of $%
\mathcal{I}$.

\subsection{The classification for the subfamily $L_{abc_{2}}$ with $c\neq 0$%
, $ab=0$}

First we show that the state $L_{abc_{2}}$ with $c\neq 0$, $b=0$ is
equivalent to the state $L_{abc_{2}}$ with $c\neq 0$, $a=0$ under SLOCC as
follows. From the state $\frac{a}{2}(|0000\rangle +|1111\rangle )-\frac{a}{2}%
(|0011\rangle +|1100\rangle )+c(\ |0101\rangle +|1010\rangle )+|0110\rangle $%
,\ we can obtain $\frac{a}{2}(|0000\rangle +|1111\rangle )+\frac{a}{2}%
(|0011\rangle +|1100\rangle )+c(\ |0101\rangle +|1010\rangle )+|0110\rangle $
under SLOCC $\alpha \otimes \beta \otimes \gamma \otimes \delta $, where $%
\alpha =\beta =diag\{i$, $1\}$, $\gamma =\delta =diag\{-i$, $1\}$.
Therefore, we only consider $L_{abc_{2}}$ with $c\neq 0$, $a=0$ below.

Furthermore, note that the state $L_{abc_{2}}$ with $c\neq 0$, $a=0$, and $%
b=-c$ is equivalent to the state $L_{abc_{2}}$ with $c\neq 0$, $a=0$, and $%
b=c$ under SLOCC. We can argue this as follows. From the state $\frac{-b}{2}%
(|0000\rangle +|1111\rangle )-\frac{-b}{2}(|0011\rangle +|1100\rangle )-b(\
|0101\rangle +|1010\rangle )+|0110\rangle $, we can obtain $\frac{b}{2}%
(|0000\rangle +|1111\rangle )-\frac{b}{2}(|0011\rangle +|1100\rangle )-b(\
|0101\rangle +|1010\rangle )+|0110\rangle $ under SLOCC $I\otimes I\otimes
\sigma _{z}\otimes (-\sigma _{z})$.

This subfamily has the following four true SLOCC entanglement classes: B5.1,
B5.2, B5.3 and B5.4.

Class B5.1, it includes states $L_{abc_{2}}$ with $c\neq 0$ and $a=b=0$.
That is, B5.1 includes $c$\ $($\ $|0101\rangle +|1010\rangle )+|0110\rangle $%
. B5.1 is a true SLOCC entanglement class because $c$\ $($\ $|0101\rangle
+|1010\rangle )+|0110\rangle =\alpha \otimes \beta \otimes \gamma \otimes
\delta (\ |0101\rangle +|1010\rangle +|0110\rangle )$, where $\alpha =\delta
=I$, $\beta =\gamma =diag(c,1)$.

Class B5.2, it includes states $L_{abc_{2}}$ with $c\neq 0$, $a=0$, and $b=c$%
. That is, B5.2 includes $\frac{b}{2}(|0000\rangle +|1111\rangle )-\frac{b}{2%
}(|0011\rangle +|1100\rangle )+b(\ |0101\rangle +|1010\rangle )+|0110\rangle 
$. B5.2 is a true SLOCC entanglement class because $L_{abc_{2}}(a=0,b=c)=%
\alpha \otimes \beta \otimes \gamma \otimes \delta L_{abc_{2}}(a=0,b=c=1)$,
where $\alpha =\delta =diag(1,\sqrt{b})$, and $\beta =\gamma =diag(\sqrt{b}%
,1)$.

Class B5.3, it includes states $L_{abc_{2}}$ with $c\neq 0$, $a=0$, $b\neq
\pm c$, and $b^{2}+2c^{2}=0$. That is, B5.3 includes $\frac{b}{2}%
(|0000\rangle +|1111\rangle )-\frac{b}{2}(|0011\rangle +|1100\rangle )\pm 
\frac{bi}{\sqrt{2}}(\ |0101\rangle +|1010\rangle )+|0110\rangle $. We can
show that B5.3 is a true SLOCC entanglement class as follows. $%
L_{abc_{2}}(a=0,c=\pm bi/\sqrt{2})=\alpha \otimes \beta \otimes \gamma
\otimes \delta L_{abc_{2}}(a=0,b=1,c=\pm i/\sqrt{2})$, where $\alpha =\delta
=diag(1,\sqrt{b})$, and $\beta =\gamma =diag(\sqrt{b},1)$. Also, $%
L_{abc_{2}}(a=0,b=1,c=i/\sqrt{2})=\alpha \otimes \beta \otimes \gamma
\otimes \delta L_{abc_{2}}(a=0,b=1,c=-i/\sqrt{2})$, where $\alpha =\sigma
_{z}$, $\beta =diag(i,1)$, $\gamma =diag(-i,1)$, and $\delta =diag(1,-i)$.

Class B5.4, it includes states $L_{abc_{2}}$ with $c\neq 0$, $a=0$, $b\neq
\pm c$, and $b^{2}+2c^{2}\neq 0$. We cannot classify B5.4 further.

Table II (4). The true SLOCC entanglement classes in the subfamily $%
L_{abc_{2}}$ with $c\neq 0$, $ab=0$

\begin{tabular}{|l|l|l|l|l|l|}
\hline
classes & criteria for classification & $\mathcal{I}$ & $D_{1}$ & $D_{2}$ & $%
D_{3}$ \\ \hline
B5.1 & $a=b=0$ & $\neq 0$ & $0$ & $\Delta $ & $0$ \\ \hline
B5.2 & $a=0$, and $b=c$ & $\neq 0$ & $\Delta $ & $\Delta $ & $\Delta $ \\ 
\hline
B5.3 & $a=0$, $b\neq \pm c$, and $b^{2}+2c^{2}=0$ & $=0$ & $\Delta $ & $%
\Delta $ & $\Delta $ \\ \hline
B5.4, * & $a=0$, $b\neq \pm c$, and $b^{2}+2c^{2}\neq 0$ & $\neq 0$ & $%
\Delta $ & $\Delta $ & $\Delta $ \\ \hline
\end{tabular}

\section{Family $L_{a_{2}b_{2}}$}

The representative state of this family is $L_{a_{2}b_{2}}=$\ $a$ $%
(|0000\rangle +|1111\rangle )+b$\ $($\ $|0101\rangle +|1010\rangle
)+|0110\rangle +|0011\rangle $. When $a=b=0$, this becomes a product state: $%
|01\rangle _{13}\otimes (|01\rangle +|10\rangle )_{24}$. We can distinguish
four true SLOCC entanglement classes in this family, which are denoted as
V1, V2, V3, and V4.

For the class V1, it includes states $L_{a_{2}b_{2}}$ with $a=\pm b\neq 0$.
V1 is a true SLOCC entanglement class.\ The representative state is denoted
as $L_{a_{2}b_{2}}(a=b=1)=$ $|0000\rangle +|1111\rangle +$\ $|0101\rangle
+|1010\rangle +|0110\rangle +|0011\rangle $. The following is our argument.
Let $\alpha =diag\{1$, $a\}$, $\gamma =\{a,1\}$, and $\beta =\delta =I$.
Then $L_{a_{2}b_{2}}(a=b\neq 0)=$ $\alpha \otimes \beta \otimes \gamma
\otimes \delta L_{a_{2}b_{2}}(a=b=1)$. Let $\alpha =diag\{-i$, $a\}$, $%
\gamma =diag\{-ai$, $1\}$, and $\beta =\delta =diag\{i$, $1\}$. Then $%
L_{a_{2}b_{2}}(a=-b\neq 0)=$ $\alpha \otimes \beta \otimes \gamma \otimes
\delta L_{a_{2}b_{2}}(a=b=1)$.

For the class V2, it includes states $L_{a_{2}b_{2}}$ with $a\neq \pm b$, $%
ab\neq 0$, and $a^{2}+b^{2}=0$. We can argue that V2\ is a true SLOCC
entanglement class as follows. Let $\alpha =diag\{1$, $a\}$, $\beta =I$, $%
\gamma =diag\{a$, $1\}$, and $\delta =I$. Then, it is easy to verify that $%
L_{a_{2}b_{2}}($with $b=\pm ai)=\alpha \otimes \beta \otimes \gamma \otimes
\delta L_{a_{2}b_{2}}(a=1$, $b=\pm i)$. Furthermore, we can demonstrate that 
$L_{a_{2}b_{2}}(a=1,b=-i)=\alpha \otimes \beta \otimes \gamma \otimes \delta
L_{a_{2}b_{2}}(a=1,b=i)$, where $\alpha =diag(1,i)$, $\beta =diag(i,1)$, $%
\gamma =diag(-i,1)$, and $\delta =diag(1,-i)$.

For the class V3, it includes states $L_{a_{2}b_{2}}$ with $a\neq \pm
b,ab\neq 0$, and $a^{2}+b^{2}\neq 0$. Each state of V3 is a true entangled
state. However we cannot classify V3 further.

For the class V4, it includes states $L_{a_{2}b_{2}}$ with $a\neq \pm b$ and 
$ab=0$. V4 is a true SLOCC entanglement class. The representative state is
denoted as $L_{a_{2}b_{2}}(a=1$, $b=0)$ $=$ $|0000\rangle +|1111\rangle
+|0110\rangle +|0011\rangle $. We can argue this as follows. We can show
that for any $a$ and $b$, $L_{a_{2}b_{2}}$ with $a\neq \pm b$ and $ab=0$ is
equivalent to the representative state $L_{a_{2}b_{2}}(a=1$, $b=0)$ under
SLOCC as follows.\ Let $\alpha =diag\{1$, $a\}$, $\gamma =diag\{a$, $1\}$, $%
\beta =\delta =I$. Then $L_{a_{2}b_{2}}(a\neq 0$, $b=0\}=\alpha \otimes
\beta \otimes \gamma \otimes \delta $ $L_{a_{2}b_{2}}(a=1$, $b=0)$. Let $%
\alpha =diag\{1$, $b\}$, $\gamma =diag\{b$, $1\}$, and $\beta =\delta
=\sigma _{x}$. Then, $L_{a_{2}b_{2}}(a=0$, $b\neq 0\}=\alpha \otimes \beta
\otimes \gamma \otimes \delta L_{a_{2}b_{2}}(a=1$, $b=0)$.

Note that the state $L_{a_{2}b_{2}}$ with $a=1$ and $b=0$, and the state $%
L_{a_{2}b_{2}}$\ with $a=0$ and $b=1$ are different representative states of
the class V4 because $|0101\rangle +|1010\rangle +|0110\rangle +|0011\rangle
=I\otimes \sigma _{x}\otimes I\otimes \sigma _{x}($ $|0000\rangle
+|1111\rangle +|0110\rangle +|0011\rangle )$.

Table III. Four true SLOCC entanglement classes in Family $L_{a_{2}b_{2}}$

\begin{tabular}{|l|l|l|l|l|l|}
\hline
classes & criteria for classification & $\mathcal{I}$ & $D_{1}$ & $D_{2}$ & $%
D_{3}$ \\ \hline
V1 & $L_{a_{2}b_{2}}$ with $a=\pm b\neq 0$ & $\neq 0$ & $\Delta $ & $0$ & $0$
\\ \hline
V2 & $L_{a_{2}b_{2}}$ with $a\neq \pm b$, $ab\neq 0$, and $a^{2}+b^{2}=0$ & $%
=0$ & $\Delta $ & $\Delta $ & $\Delta $ \\ \hline
V3, * & $L_{a_{2}b_{2}}$ with $a\neq \pm b$, $ab\neq 0$, and $%
a^{2}+b^{2}\neq 0$ & $\neq 0$ & $\Delta $ & $\Delta $ & $\Delta $ \\ \hline
V4 & $L_{a_{2}b_{2}}$ with $a\neq \pm b$ and $ab=0\ $ & $\neq 0$ & $\Delta $
& $\Delta $ & $\Delta $ \\ \hline
\end{tabular}

To demonstrate that the four classes in Table III are distinct true SLOCC
entanglement classes, we only need to show that the class V4 is different
from the class V3\ due to the properties of $\mathcal{I}$,$\ D_{1}$, $D_{2}$%
, and $D_{3}$. For the class V4, each state of the class V4 has the
following $F_{i}$:

$F_{1}=\allowbreak a^{4}\alpha _{1}^{2}\alpha _{2}^{2}P$, $F_{2}=\allowbreak
a^{4}\alpha _{3}^{2}\alpha _{4}^{2}P$, $F_{3}=\allowbreak a^{4}\beta
_{1}^{2}\beta _{2}^{2}Q$, $F_{4}=\allowbreak a^{4}\allowbreak \beta
_{3}^{2}\beta _{4}^{2}Q$, $F_{5}=\allowbreak a^{4}\gamma _{2}^{2}\gamma
_{1}^{2}R$, $F_{6}=\allowbreak a^{4}\allowbreak \gamma _{4}^{2}\gamma
_{3}^{2}R$, $F_{7}=\allowbreak a^{4}\allowbreak \delta _{1}^{2}\delta
_{2}^{2}S$, $F_{8}=\allowbreak a^{4}\delta _{3}^{2}\delta _{4}^{2}S$,

$F_{9}=\allowbreak a^{4}\allowbreak \left( \alpha _{1}\alpha _{4}\beta
_{1}\beta _{4}-\alpha _{2}\alpha _{3}\beta _{2}\beta _{3}\right)
^{2}\det^{2}(\delta )\det^{2}(\gamma )$,

$F_{10}=\allowbreak a^{4}\left( -\alpha _{1}\alpha _{4}\beta _{2}\beta
_{3}+\alpha _{2}\alpha _{3}\beta _{1}\beta _{4}\right) ^{2}\allowbreak
\det^{2}(\delta )\det^{2}(\gamma )$.

Clearly, for each state of the class V4, the above $F_{i}$ satisfy
Properties 2.1 and 2.2 in Appendix A. However, in the class V3 the state $%
L_{a_{2}b_{2}}$ with $a\neq \pm b$, $ab\neq 0$, and $a^{2}+b^{2}\neq 0$
satisfies the following

\begin{eqnarray}
F_{i} &=&0\text{, }i=2\text{, }3\text{, }4\text{, }5\text{, }7\text{, }8;
\label{V4-F-1} \\
F_{1} &=&F_{6}=4ab\text{, }F_{9}=a^{4}\text{, }F_{10}=b^{4}.  \label{V4-F-2}
\end{eqnarray}

Clearly, these $F_{i}$ in Eqs. (\ref{V4-F-1}) and Eq. (\ref{V4-F-2}) do not
satisfy property 2.2 in appendix A. Therefore the class V4 is different from
the class V3.

\section{Family $L_{ab_{3}}$}

The representative state of this family is $L_{ab_{3}}=a$ $(|0000\rangle
+|1111\rangle )+\frac{a+b}{2}($\ $|0101\rangle +|1010\rangle )+$\ $\frac{a-b%
}{2}($\ $|0110\rangle +|1001\rangle )+\frac{i}{\sqrt{2}}(|0001\rangle
+|0010\rangle +|0111\rangle +|1011\rangle )$.

Let us consider three subfamilies, which are the subfamily $L_{ab_{3}}$ with 
$a=\pm b$, the subfamily $L_{ab_{3}}$ with $a\neq \pm b$ and $ab=0$, and the
subfamily $L_{ab_{3}}$ with $a\neq \pm b$ and $ab\neq 0$. We demonstrate
that these three subfamilies are inequivalent under SLOCC and there are at
least eight true SLOCC entanglement classes in this family below.

The state $L_{ab_{3}}$ satisfies

$\mathcal{I}=-(b^{2}+3a^{2})$, $D_{1}=a^{2}(a+b)^{2}/4$, $%
D_{2}=(a^{2}-b^{2})^{2}/16$, $D_{3}=-a^{2}(a-b)^{2}/4$,

\begin{eqnarray}
F_{1} &=&F_{3}=F_{6}=F_{8}=(a^{2}-b^{2})/2,  \notag \\
F_{2} &=&F_{4}=F_{5}=F_{7}=0,  \notag \\
F_{9} &=&a^{4},F_{10}=a^{2}b^{2}.  \label{Lab3-state-F}
\end{eqnarray}

Let $|\psi \rangle $ in Eq. (\ref{SLOCC-1}) be the state $L_{ab_{3}}$. Let $%
|\psi ^{\prime }\rangle $ be any state which is equivalent to $|\psi \rangle 
$\ under SLOCC. By solving matrix equation in Eq. (\ref{SLOCC-1}), we obtain
the amplitudes $a_{i}$ of the state $|\psi ^{\prime }\rangle $. By
substituting $a_{i}$ into $F_{i}$, we obtain the following $F_{i}$.\ That
is, each state in Family $L_{ab_{3}}$ satisfies

\begin{eqnarray}
F_{1} &=&(a^{2}-b^{2})\alpha _{1}^{4}P/2,F_{2}=(a^{2}-b^{2})\alpha
_{3}^{4}P/2,  \notag \\
F_{3} &=&(a^{2}-b^{2})\beta _{1}^{4}Q/2,F_{4}=(a^{2}-b^{2})\beta _{3}^{4}Q/2,
\notag \\
F_{5} &=&(a^{2}-b^{2})\gamma _{2}^{4}R/2,F_{6}=(a^{2}-b^{2})\gamma
_{4}^{4}R/2,  \notag \\
F_{7} &=&(a^{2}-b^{2})\delta _{2}^{4}S/2,F_{8}=(a^{2}-b^{2})\delta
_{4}^{4}S/2.  \label{Lab3-class-F}
\end{eqnarray}

\subsection{Subfamily $L_{ab_{3}}$ with $a=\pm b$}

This subfamily has the following three classes: R1.1, R1.2 and R1.3.

For the class R1.1, it includes the state $L_{ab_{3}}$ with $a=b=0$.\ Let $%
L_{ab_{3}}(a=b=0)=|0001\rangle +|0010\rangle +|0111\rangle +|1011\rangle $.
Clearly, $L_{ab_{3}}(a=b=0)$ is equivalent to the state $|W\rangle $ because 
$L_{ab_{3}}(a=b=0)=I\otimes I\otimes \sigma _{x}\otimes \sigma _{x}|W\rangle 
$.

For the class R1.2, it includes states $L_{ab_{3}}$ with $a=b\neq 0$. That
is, R1.2 includes $L_{ab_{3}}(a=b)=a(|0000\rangle +|1111\rangle +$\ $%
|0101\rangle +|1010\rangle )+\frac{i}{\sqrt{2}}(|0001\rangle +|0010\rangle
+|0111\rangle +|1011\rangle )$. A representative state is $%
L_{ab_{3}}(a=b=1)=|0000\rangle +|1111\rangle +$\ $|0101\rangle +|1010\rangle
+\frac{i}{\sqrt{2}}(|0001\rangle +|0010\rangle +|0111\rangle +|1011\rangle )$%
. R1.2 is a true SLOCC entanglement class because $L_{ab_{3}}(a=b)=\alpha
\otimes \beta \otimes \gamma \otimes \delta L_{ab_{3}}(a=b=1)$, where $%
\alpha =\beta =diag\{1$, $a\}$, $\gamma =diag\{a$, $1\}$, and $\delta
=diag\{1$, $1/a\}$.

For the class R1.3, it includes states $L_{ab_{3}}$ with $a=-b\neq 0$. That
is, R1.3 includes $L_{ab_{3}}(a=-b)=a(|0000\rangle +|1111\rangle +$\ $%
|0110\rangle +|1001\rangle )+\frac{i}{\sqrt{2}}(|0001\rangle +|0010\rangle
+|0111\rangle +|1011\rangle )$.\ A representative state is $%
L_{ab_{3}}(a=-b=1)=$ $|0000\rangle +|1111\rangle +$\ \ $|0110\rangle
+|1001\rangle +\frac{i}{\sqrt{2}}(|0001\rangle +|0010\rangle +|0111\rangle
+|1011\rangle )$. R1.3 is a true SLOCC entanglement class because $%
L_{ab_{3}}(a=-b)=\alpha \otimes \beta \otimes \gamma \otimes \delta
L_{ab_{3}}(a=-b=1)$, where $\alpha =\beta =diag\{1$, $a\}$, $\gamma =diag\{a$%
, $1\}$, and $\delta =diag\{1$, $1/a\}$.

Classes R1.1, R1.2, and R1.3 are different under SLOCC by the values of $%
\mathcal{I}$,$\ D_{1}$, $D_{2}$, and $D_{3}$ in Table IV.

\subsection{Subfamily $L_{ab_{3}}$ with $a\neq \pm b$ and $ab=0$}

Subfamily $L_{ab_{3}}$ with $a\neq \pm b$ and $ab=0$ has two classes denoted
as R2.1 and R2.2.

For the class R2.1, it includes states $L_{ab_{3}}$ with $a=0,b\neq 0$. That
is, R2.1 includes $L_{ab_{3}}(a=0,b\neq 0)=\frac{b}{2}($\ $|0101\rangle
+|1010\rangle )-$\ $\frac{b}{2}($\ $|0110\rangle +|1001\rangle )+\frac{i}{%
\sqrt{2}}(|0001\rangle +|0010\rangle +|0111\rangle +|1011\rangle )$. R2.1 is
a true entanglement class because $L_{ab_{3}}(a=0,b\neq 0)=\alpha \otimes
\beta \otimes \gamma \otimes \delta L_{ab_{3}}(a=0,b=1)$, where $\alpha
=\beta =diag\{1$, $b\}$, $\gamma =diag\{b$, $1\}$, and $\delta =diag\{1$, $%
1/b\}$.

For the class R2.2, it includes states $L_{ab_{3}}$ with $a\neq 0,b=0$. That
is, R2.2 includes $L_{ab_{3}}(b=0)=a$ $(|0000\rangle +|1111\rangle )+\frac{a%
}{2}($\ $|0101\rangle +|1010\rangle )+$\ $\frac{a}{2}($\ $|0110\rangle
+|1001\rangle )+\frac{i}{\sqrt{2}}(|0001\rangle +|0010\rangle +|0111\rangle
+|1011\rangle )$. R2.2 is a true entanglement class because $%
L_{ab_{3}}(b=0)=\alpha \otimes \beta \otimes \gamma \otimes \delta
L_{ab_{3}}(a=1,b=0)$, where $\alpha =\beta =diag\{1$, $a\}$, $\gamma
=diag\{a $, $1\}$, and $\delta =diag\{1$, $1/a\}$.

We argue that classes R2.1 and R2.2 are different under SLOCC\ as follows.

For each state of the class R2.1, $F_{i}$ ($i=1$, ..., $8$) can be obtained
from Eq. (\ref{Lab3-class-F}) by letting $a=0$,

$F_{9}=-\frac{1}{2}b^{2}(\alpha _{2}\alpha _{3}\beta _{1}\beta _{3}-\alpha
_{1}\alpha _{4}\beta _{1}\beta _{3}-\alpha _{1}\alpha _{3}\beta _{2}\beta
_{3}+\alpha _{1}\alpha _{3}\beta _{1}\beta _{4})^{2}\det^{2}(\gamma
)\det^{2}(\delta )$,

$F_{10}=-\frac{1}{2}b^{2}(\alpha _{2}\alpha _{3}\beta _{1}\beta _{3}-\alpha
_{1}\alpha _{4}\beta _{1}\beta _{3}+\alpha _{1}\alpha _{3}\beta _{2}\beta
_{3}-\alpha _{1}\alpha _{3}\beta _{1}\beta _{4})^{2}\det^{2}(\gamma
)\det^{2}(\delta )$.

Then, we derive that for each state in the class R2.1, $F_{i}$ satisfy the
following:

Property 7.1.\textbf{\ }If $F_{i}=0$ then $F_{9}=F_{10}$, $i=1$, $2$, $3$,
or $4$.

Property 7.2. If ${\small F_{1}F_{2}=0}\wedge {\small F}_{3}{\small F}_{4}=%
{\small 0}$ then ${\small F}_{9}{\small =F}_{10}{\small =0}.$

For each state of the class R2.2, $F_{i}$ ($i=1$, ..., $8$) can be obtained
from Eq. (\ref{Lab3-class-F}) by letting $b=0$. $F_{9}$ and $F_{10}$ are
omitted because they are too complicated. We can derive that for each state
of the class R2.2, $F_{i}$\ satisfy the following properties:

Property 8.1.\textbf{\ }If $F_{1}=F_{3}=0$ or $F_{2}=F_{4}=0$\ then $%
F_{9}\neq 0\wedge F_{10}=0$.

Property 8.2\textbf{. }If $F_{1}=F_{4}=0$ or $F_{2}=F_{3}=0$\ then $%
F_{9}=0\wedge F_{10}\neq 0.$

Classes R2.1 and R2.2 are different under SLOCC by the following argument.

From Eq. (\ref{Lab3-state-F}),\ the state $L_{ab_{3}}(b=0)$\ in the class
R2.2 satisfies $F_{1}=F_{3}=F_{6}=F_{8}=a^{2}/2$, $F_{2}=F_{4}=F_{5}=F_{7}=0$%
, $F_{9}=a^{4}$, $F_{10}=0$. It is obvious that the state $L_{ab_{3}}(b=0)$
in the class R2.2 does not satisfy property 7.2. Therefore, classes R2.1 and
R2.2 are different under SLOCC.

\subsection{Subfamily $L_{ab_{3}}$ with $a\neq \pm b$ and $ab\neq 0$}

This subfamily has two subsubfamilies denoted as R3.1 and R3.2.\ 

Subsubfamily R3.1 includes states $L_{ab_{3}}$ with $a\neq \pm b,ab\neq
0,3a^{2}+b^{2}\neq 0$. Each one of this subsubfamily is a true entanglement
state. However, we cannot classify R3.1 further.

Subsubfamily R3.2 includes states $L_{ab_{3}}$ with $a\neq \pm b,ab\neq
0,3a^{2}+b^{2}=0$. This subsubfamily\ consists of two inequivalent true
SLOCC entanglement classes. Their representative states are denoted as $%
L_{ab_{3}}(a=1$, $b=\pm \sqrt{3}i)$ $=$ $|0000\rangle +|1111\rangle +\frac{%
(1\pm \sqrt{3}i)}{2}($\ $|0101\rangle +|1010\rangle )+$\ $\frac{(1\mp \sqrt{3%
}i)}{2}($\ $|0110\rangle +|1001\rangle )+\frac{i}{\sqrt{2}}(|0001\rangle
+|0010\rangle +|0111\rangle +|1011\rangle )$. We can argue this as follows.
Let $\alpha =\beta =diag\{1$, $a\}$, $\gamma =diag\{a$, $1\}$, and $\delta
=diag\{1$, $1/a\}$. Then, it is easy to verify that the state $L_{ab_{3}}$
(with $a\neq \pm b$, $ab\neq 0$ and $b=\pm \sqrt{3}ai$) $=\alpha \otimes
\beta \otimes \gamma \otimes \delta L_{ab_{3}}(a=1$, $b=\pm \sqrt{3}i)$.
Furthermore, we can demonstrate that these two representative states are
inequivalent under SLOCC.

Subfamilies R3.1 and R3.2 are different under SLOCC because they have
different values of $\mathcal{I}$ in Table IV.\ 

\subsection{Inequivalent subfamilies}

Lemma 6.1. The subfamily $L_{ab_{3}}$ with $a\neq \pm b$ and $ab=0$, and the
subfamily $L_{ab_{3}}$ with $a\neq \pm b$ and $ab\neq 0$ are inequivalent to
the subfamily $L_{ab_{3}}$ with $a=\pm b$ under SLOCC.

Proof. From Eq. (\ref{Lab3-class-F}), it is easy to see that $F_{i}=0$, $i=1$%
, ..., $8$, for any state in the subfamily $L_{ab_{3}}$ with $a=\pm b$.
While $F_{i}\neq 0$, $i=1$, ..., $8$, for the state $L_{ab_{3}}$ with $a\neq
\pm b$ and $ab=0$ and the state $L_{ab_{3}}$ with $a\neq \pm b$ and $ab\neq
0 $. Hence, the subfamily $L_{ab_{3}}$ with $a\neq \pm b$ and $ab=0$ and the
subfamily $L_{ab_{3}}$ with $a\neq \pm b$ and $ab\neq 0$ are inequivalent to
the subfamily $L_{ab_{3}}$ with $a=\pm b$ under SLOCC, respectively.

Lemma 6.2. The subfamily $L_{ab_{3}}$ with $a\neq \pm b$ and $ab=0$, and the
subfamily $L_{ab_{3}}$ with $a\neq \pm b$ and $ab\neq 0$ are inequivalent
under SLOCC.

Proof.

When $a\neq \pm b$ and $ab\neq 0$, $F_{i}$\ in\ Eq. (\ref{Lab3-state-F}) do
not satisfy properties 7.2 or 8.1. Hence, lemma 6.2 holds.

Conclusively, Family $L_{ab_{3}}$ has\ at least eight distinct true
entanglement classes. See Table IV. It means that Family $L_{ab_{3}}$ does
not include any product state.

Table IV. Eight True SLOCC entanglement classes in Family \ $L_{ab_{3}}$\ \
\ 

\begin{tabular}{|l|l|l|l|l|l|}
\hline
classes & criteria for classification & $\mathcal{I}$ & $D_{1}$ & $D_{2}$ & $%
D_{3}$ \\ \hline
R1.1 (i.e., $|W\rangle $) & $L_{ab_{3}}$ with $a=b=0$ & $=0$ & $0$ & $0$ & $%
0 $ \\ \hline
R1.2 & $L_{ab_{3}}$ with $a=b\neq 0$ & $\neq 0$ & $\Delta $ & $0$ & $0$ \\ 
\hline
R1.3 & $L_{ab_{3}}$ with $a=-b\neq 0$ & $\neq 0$ & $0$ & $0$ & $\Delta $ \\ 
\hline
R2.1 & $L_{ab_{3}}$ with $a=0,b\neq 0$ & $\neq 0$ & $\Delta $ & $\Delta $ & $%
\Delta $ \\ \hline
R2.2 & $L_{ab_{3}}$ with $a\neq 0,b=0$ & $\neq 0$ & $\Delta $ & $\Delta $ & $%
\Delta $ \\ \hline
R3.1, * & $L_{ab_{3}}$ with $a\neq \pm b,ab\neq 0,3a^{2}+b^{2}\neq 0$ & $%
\neq 0$ & $\Delta $ & $\Delta $ & $\Delta $ \\ \hline
subsubfamily R3.2 (two classes) & $L_{ab_{3}}$ with $a\neq \pm b,ab\neq
0,3a^{2}+b^{2}=0$ & $=0$ & $\Delta $ & $\Delta $ & $\Delta $ \\ \hline
\end{tabular}

\section{Other five families}

\subsection{Family $L_{a_{4}}$}

The representative state of this family is $L_{a_{4}}=a(|0000\rangle
+|0101\rangle +|1010\rangle +|1111\rangle )+(i|0001\rangle +|0110\rangle
-i|1011\rangle )$. There are two true SLOCC entanglement classes in this
family.

(1). The class $L_{a_{4}}$ with $a=0$. In this case, $L_{a_{4}}$ reduces to $%
i|0001\rangle +|0110\rangle -i|1011\rangle $. Clearly, this is a true
entanglement\ class.

(2). The class $L_{a_{4}}$ with $a\neq 0$. We show that this is a true SLOCC
entanglement class as follows. Let $L_{a_{4}}(a=1)=|0000\rangle
+|0101\rangle +|1010\rangle +|1111\rangle +(i|0001\rangle +|0110\rangle
-i|1011\rangle )$. Let $\alpha =diag\{1$, $a^{2}\}$, $\beta =diag\{1$, $a\}$%
; $\gamma =diag\{1$, $1/a^{2}\}$, and $\delta =diag\{a$, $1\}$. Then, it is
easy to verify that for any $a\neq 0$, $L_{a_{4}}=\alpha \otimes \beta
\otimes \gamma \otimes \delta L_{a_{4}}(a=1)$.

Therefore, there are two true SLOCC entanglement classes in Family \ $%
L_{a_{4}}$ and these two classes are different because they have different
values of $\mathcal{I}$. See Table V.

For the class $L_{a_{4}}$ with $a\neq 0$, $\mathcal{I}=2a^{2}T$.\ For $SL$%
-operations, $\mathcal{I}=2a^{2}$.\ It means that the SLOCC entanglement
class $L_{a_{4}}$ with $a\neq 0$\ includes a continuous parameter of $SL$
classes. In other words,\ the class $L_{a_{4}}$ with $a\neq 0$\ can be
described by a continuous parameter.

Table V. Two true SLOCC entanglement classes in Family \ $L_{a_{4}}$

\begin{tabular}{|l|l|l|l|l|}
\hline
& $\mathcal{I}$ & $D_{1}$ & $D_{2}$ & $D_{3}$ \\ \hline
$L_{a_{4}}$ with $a=0$ & $0$ & $\Delta $ & $0$ & $0$ \\ \hline
$L_{a_{4}}$ with $a\neq 0$ & $\neq 0$ & $\Delta $ & $\Delta $ & $\Delta $ \\ 
\hline
\end{tabular}

\subsection{Family $L_{a_{2}0_{3\oplus 1}}$}

The representative state of this family is $L_{a_{2}0_{3\oplus 1}}=$ $%
a(|0000\rangle +|1111\rangle )+|0011\rangle +|0101\rangle +|0110\rangle $.
We argue there are only two SLOCC entanglement classes in this family as
follows.

(1). The class $L_{a_{2}0_{3\oplus 1}}$ with $a=0$. In this case, $%
L_{a_{2}0_{3\oplus 1}}$ becomes $|0\rangle \otimes (|011\rangle +|101\rangle
+|110\rangle )$, which is a product state of the one-qubit state $|0\rangle $%
\ and the 3-qubit $|W\rangle $.

(2). The class $L_{a_{2}0_{3\oplus 1}}$ with $a\neq 0$.\ We show that this
is a true SLOCC entanglement class as follows.

The state $L_{a_{2}0_{3\oplus 1}}$ with $a\neq 0$ is a true entanglement
state. Let $L_{a_{2}0_{3\oplus 1}}(a=1)=$ $|0000\rangle +|1111\rangle
+|0011\rangle +|0101\rangle +|0110\rangle $. Let $\alpha =diag\{\sqrt{a}$, $%
a^{2}\}$, $\beta =diag\{1/\sqrt{a}$, $1/a\}$, $\gamma =diag\{\sqrt{a}$, $1\}$%
, and $\delta =diag\{\sqrt{a}$, $1\}$. Then it is easy to verify that $%
L_{a_{2}0_{3\oplus 1}}(a\neq 0)=\alpha \otimes \beta \otimes \gamma \otimes
\delta L_{a_{2}0_{3\oplus 1}}(a=1)$.

For the class $L_{a_{2}0_{3\oplus 1}}$ with $a\neq 0$, $\mathcal{I}=a^{2}T$%
.\ For $SL$-operations, $\mathcal{I}=a^{2}$.\ It implies that the SLOCC
entanglement class $L_{a_{2}0_{3\oplus 1}}$ with $a\neq 0$\ includes a
continuous parameter of $SL$ classes. It says that\ the class $%
L_{a_{2}0_{3\oplus 1}}$ with $a\neq 0$\ can be characterized by a continuous
parameter.

Table VI. One true SLOCC entanglement class in Family \ $L_{a_{2}0_{3\oplus
1}}$

\begin{tabular}{|l|l|l|l|l|}
\hline
criterion for classification & $\mathcal{I}$ & $D_{1}$ & $D_{2}$ & $D_{3}$
\\ \hline
$L_{a_{2}0_{3\oplus 1}}$ with $a\neq 0$ & $\neq 0$ & $\Delta $ & $\Delta $ & 
$\Delta $ \\ \hline
\end{tabular}

\subsection{Family $L_{0_{5\oplus 3}}$}

The representative state of this family is $L_{0_{5\oplus 3}}=|0000\rangle
+|0101\rangle +|1000\rangle +|1110\rangle $. This family is a true SLOCC
entanglement class. Each state of this family satisfies the following:

$\mathcal{I}=0$, $D_{1}=D_{2}=D_{3}=0$, $F_{1}=F_{2}=0$, $\left\vert
F_{3}\right\vert +\left\vert F_{4}\right\vert \neq 0$, $\left\vert
F_{5}\right\vert +\left\vert F_{6}\right\vert \neq 0$, $\left\vert
F_{7}\right\vert +\left\vert F_{8}\right\vert \neq 0$, $F_{9}=F_{10}$, $%
F_{3}F_{4}=(F_{9})^{2}$.

\subsection{Family $L_{0_{7\oplus 1}}$}

The representative state of this family is $L_{0_{7\oplus 1}}=|0000\rangle
+|1011\rangle +|1101\rangle +|1110\rangle $. This family is a true SLOCC
entanglement class. Each state of this family satisfies the following:

$\mathcal{I}=0$, $D_{1}$ is $\Delta $, $D_{2}$ is $\Delta $,$\ D_{3}$ is $%
\Delta $, $\left\vert F_{3}\right\vert +\left\vert F_{4}\right\vert \neq 0$, 
$\left\vert F_{5}\right\vert +\left\vert F_{6}\right\vert \neq 0$, $%
\left\vert F_{7}\right\vert +\left\vert F_{8}\right\vert \neq 0$.

\subsection{Family $L_{0_{3+\bar{1}}0_{3+\bar{1}}}$}

The representative state of this family is $L_{0_{3+\bar{1}}0_{3+\bar{1}%
}}=|0\rangle (|000\rangle +|111\rangle )$, which is a product state of the
one-qubit state $|0\rangle $ and the three-qubit $|GHZ\rangle $ state.

\textbf{Summary}

Verstraete, Dehaene, and Verschelde proposed nine families of states
corresponding to nine different ways of entangling four qubits \cite{Moor2}.
In this paper, we investigate SLOCC classification of each of the nine
families, and distinguish 49 true SLOCC entanglement classes from them. We
give complete SLOCC classifications for Families $L_{a_{4}}$, $%
L_{a_{2}0_{3\oplus 1}}$, $L_{0_{5\oplus 3}}$, $L_{0_{7\oplus 1}}$ and $%
L_{0_{3+\bar{1}}0_{3+\bar{1}}}$. But we cannot guarantee that SLOCC
classifications\ for Families $G_{abcd}$, $L_{abc_{2}}$, $L_{a_{2}b_{2}}$
and $L_{ab_{3}}$ are complete.

\section{Appendix A: Classification for Family $L_{abc_{2}}$}

\setcounter{equation}{0}\renewcommand{\theequation}{A\arabic{equation}}

The representative state of Family $L_{abc_{2}}$ satisfies the following:

$\mathcal{I}=-(a^{2}+b^{2}+2c^{2})/2$,

\begin{eqnarray}
D_{1} &=&(a+b)^{2}c^{2}/4,D_{2}=-(a^{2}-b^{2})^{2}/16,D_{3}=(a-b)^{2}c/4, 
\notag \\
F_{1} &=&F_{4}=F_{6}=F_{7}=(a^{2}-b^{2})c,  \notag \\
F_{2} &=&F_{3}=F_{5}=F_{8}=0,F_{9}=a^{2}b^{2},F_{10}=c^{4}.
\label{Labc2-state-F}
\end{eqnarray}

\subsection{Subfamily $L_{abc_{2}}$\ with $c=0$.}

$L_{abc_{2}}$ becomes $L_{abc_{2}}(c=0)=\frac{a+b}{2}(|0000\rangle
+|1111\rangle )+\frac{a-b}{2}(|0011\rangle +|1100\rangle )+|0110\rangle $.
Each state of this subfamily satisfies the following:

$\mathcal{I}=\frac{1}{2}(a^{2}+b^{2})T$,

$D_{1}=\frac{1}{2}ab(a-b)\alpha _{1}\alpha _{3}\gamma _{2}\gamma _{4}\det
(\alpha )\det^{2}(\beta )\det (\gamma )\det^{2}(\delta )$,

$D_{2}=\frac{1}{16}(a^{2}-b^{2})\det (\alpha )\det (\beta )\det^{2}(\gamma
)\det^{2}(\delta )$

$(2a^{2}\alpha _{1}\alpha _{3}\beta _{1}\beta _{3}+2b^{2}\alpha _{1}\alpha
_{3}\beta _{1}\beta _{3}+a^{2}\alpha _{2}\alpha _{3}\beta _{2}\beta
_{3}-b^{2}\alpha _{2}\alpha _{3}\beta _{2}\beta _{3}+a^{2}\alpha _{1}\alpha
_{4}\beta _{2}\beta _{3}-b^{2}\alpha _{1}\alpha _{4}\beta _{2}\beta _{3}$

$+a^{2}\alpha _{2}\alpha _{3}\beta _{1}\beta _{4}-b^{2}\alpha _{2}\alpha
_{3}\beta _{1}\beta _{4}+a^{2}\alpha _{1}\alpha _{4}\beta _{1}\beta
_{4}-b^{2}\alpha _{1}\alpha _{4}\beta _{1}\beta _{4}+2a^{2}\alpha _{2}\alpha
_{4}\beta _{2}\beta _{4}+2b^{2}\alpha _{2}\alpha _{4}\beta _{2}\beta _{4})$

$D_{3}=-\frac{1}{2}ab(a+b)\alpha _{1}\alpha _{3}\delta _{1}\delta _{3}\det
(\alpha )\det^{2}(\beta )\det^{2}(\gamma )\det (\delta )$, 
\begin{eqnarray}
F_{1} &=&a^{2}b^{2}\alpha _{1}^{2}\alpha _{2}^{2}P,F_{2}=a^{2}b^{2}\alpha
_{3}^{2}\alpha _{4}^{2}P,F_{3}=a^{2}b^{2}\allowbreak \beta _{1}^{2}\beta
_{2}^{2}Q,F_{4}=a^{2}b^{2}\allowbreak \beta _{3}^{2}\beta _{4}^{2}Q,  \notag
\\
F_{5} &=&a^{2}b^{2}\allowbreak \gamma _{2}^{2}\gamma
_{1}^{2}R,F_{6}=a^{2}b^{2}\allowbreak \gamma _{4}^{2}\gamma
_{3}^{2}R,F_{7}=a^{2}b^{2}\allowbreak \delta _{1}^{2}\delta
_{2}^{2}S,F_{8}=\allowbreak a^{2}b^{2}\delta _{3}^{2}\delta _{4}^{2}S, 
\notag \\
F_{9} &=&a^{2}b^{2}(\alpha _{2}\alpha _{3}\beta _{2}\beta _{3}-\alpha
_{1}\alpha _{4}\beta _{1}\beta _{4})^{2}\det^{2}(\gamma )\det^{2}(\delta ), 
\notag \\
F_{10} &=&a^{2}b^{2}(\alpha _{1}\alpha _{4}\beta _{2}\beta _{3}-\alpha
_{2}\alpha _{3}\beta _{1}\beta _{4})^{2}\det^{2}(\gamma )\det^{2}(\delta ).
\label{Labc2-Fi-1}
\end{eqnarray}%
From Eq. (\ref{Labc2-Fi-1}), for each state of this subfamily $L_{abc_{2}}$
with $c=0$, $F_{i}$ have the following properties.

Property 2.1.\textbf{\ }If $ab\neq 0$ and $F_{i}=0$ then $\left\vert
F_{9}\right\vert +\left\vert F_{10}\right\vert \neq 0$, $i=1$, $2$, $3$, or $%
4$.

Property 2.2.\textbf{\ }If $ab\neq 0$ and ${\small F_{1}F_{2}=F}_{3}{\small F%
}_{4}={\small 0}$ then ${\small F}_{9}{\small =0}\wedge {\small F}_{10}%
{\small \neq 0}$ or ${\small F}_{9}{\small \neq 0}\wedge {\small F}_{10}%
{\small =0}.$

Property 2.3. If $ab\neq 0$, $a\neq \pm b$, and ${\small F_{1}=F_{2}=F}_{3}=%
{\small F}_{4}={\small 0}$, then $D_{2}\neq 0$.

Property 2.4. If ${\small F_{1}=F_{2}=0}$ or ${\small F}_{5}={\small F}_{6}=%
{\small 0}$ then $D_{1}=0$.

Property 2.5. If ${\small F_{1}=F_{2}=0}$ or ${\small F}_{7}={\small F}_{8}=%
{\small 0}$ then $D_{3}=0$.

From the Table II (1), it is not hard to see that the classes B1.1, B1.2,
B1.3, B1.4, and B1.5 are different from each other by the values of $%
\mathcal{I}$, $D_{1}$, $D_{2}$, $D_{3}$. Note that the state $L_{abc_{2}}$
with $c=a=0\wedge b\neq 0$ and the state $L_{abc_{2}}$\ with $c=b=0\wedge
a\neq 0$\ are different representatives of the class B1.5 because these two
states are obtained from each other by SLOCC $I\otimes \sigma _{z}\otimes
\sigma _{z}\otimes I$.

\subsection{Subfamily $L_{abc_{2}}$\ with $abc\neq 0$}

This subfamily is different from the subfamily $L_{abc_{2}}$ with $c=0$
under SLOCC. This is because from Eq. (\ref{Labc2-state-F}), it is
straightforward that for the representative state $L_{abc_{2}}$ with $%
abc\neq 0$, $F_{i}$ do not satisfy property 2.2. It implies that classes
B2.1-B2.3, B3.1-B3.3, B4.1 and B4.2 in Table II (2) are different from
classes B1.1-B1.5 in Table II (1).

\subsubsection{Subsubfamily $L_{abc_{2}}$\ with $abc\neq 0$ and $a=b$}

$L_{abc_{2}}$ becomes $L_{abc_{2}}(a=b)=$\ $a$ $(|0000\rangle +|1111\rangle
)+c$\ $($\ $|0101\rangle +|1010\rangle )+|0110\rangle $. For any state which
is connected with $L_{abc_{2}}(a=b)$ by SLOCC, we obtain the following $%
F_{i} $ and $D_{i}$.

\begin{eqnarray}
F_{1} &=&\allowbreak (a^{2}-c^{2})^{2}\alpha _{1}^{2}\alpha
_{2}^{2}P,F_{2}=\allowbreak (a^{2}-c^{2})^{2}\alpha _{3}^{2}\alpha _{4}^{2}P,
\notag \\
F_{3} &=&\allowbreak (a^{2}-c^{2})^{2}\beta _{1}^{2}\beta
_{2}^{2}Q,F_{4}=\allowbreak (a^{2}-c^{2})^{2}\allowbreak \beta _{3}^{2}\beta
_{4}^{2}Q,  \notag \\
F_{5} &=&\allowbreak (a^{2}-c^{2})^{2}\gamma _{2}^{2}\gamma
_{1}^{2}R,F_{6}=\allowbreak (a^{2}-c^{2})^{2}\allowbreak \gamma
_{4}^{2}\gamma _{3}^{2}R,  \notag \\
F_{7} &=&\allowbreak (a^{2}-c^{2})^{2}\allowbreak \delta _{1}^{2}\delta
_{2}^{2}S,F_{8}=\allowbreak (a^{2}-c^{2})^{2}\delta _{3}^{2}\delta _{4}^{2}S.
\label{Labc2-F-2}
\end{eqnarray}

$D_{1}=ac\det (\alpha )\det^{2}(\beta )\det (\gamma )\det^{2}(\delta
)(a^{2}\alpha _{1}\alpha _{3}\gamma _{1}\gamma _{3}+c^{2}\alpha _{1}\alpha
_{3}\gamma _{1}\gamma _{3}+ac\alpha _{2}\alpha _{3}\gamma _{2}\gamma
_{3}+ac\alpha _{1}\alpha _{4}\gamma _{2}\gamma _{3}+ac\alpha _{2}\alpha
_{3}\gamma _{1}\gamma _{4}+ac\alpha _{1}\alpha _{4}\gamma _{1}\gamma
_{4}+a^{2}\alpha _{2}\alpha _{4}\gamma _{2}\gamma _{4}+c^{2}\alpha
_{2}\alpha _{4}\gamma _{2}\gamma _{4})$,

$D_{2}=-c(a^{2}-c^{2})\alpha _{1}\alpha _{3}\beta _{2}\beta _{4}\det (\alpha
)\det (\beta )\det^{2}(\gamma )\det^{2}(\delta )$,

$D_{3}=-a(a^{2}-c^{2})\alpha _{1}\alpha _{3}\delta _{1}\delta _{3}\det
(\alpha )\det^{2}(\beta )\det^{2}(\gamma )\det (\delta )$,

When $a\neq \pm c$, from Eq. (\ref{Labc2-F-2}) and the above $D_{1}$, $D_{2}$
and $D_{3}$, it can be verified that the following properties 3.1-3.5 hold
for each state which is connected with $L_{abc_{2}}$ with $abc\neq 0$, $a=b$
and $a\neq \pm c$ by SLOCC.

Property 3.1. If $F_{1}=F_{2}=0$, then $D_{2}=D_{3}=0$.

Property 3.2. If $F_{3}=F_{4}=0$, then $D_{2}=0$.

Property 3.3\textbf{.} If $F_{7}=F_{8}=0$, then $D_{3}=0$.

Property 3.4\textbf{.} If $F_{1}=F_{2}=F_{5}=F_{6}=0$, then $D_{1}\neq 0$.

Property 3.5. If $F_{1}=F_{2}=F_{3}=F_{4}=0$, then $F_{9}\neq 0$ and $%
F_{10}\neq 0.$

We argue that property 3.4 holds as follows.

When $F_{1}=F_{2}=0$ and $F_{5}=F_{6}=0$, there are following four cases:
case 1. $\alpha _{1}=\alpha _{4}=0$ and $\gamma _{1}=\gamma _{4}=0$, case 2. 
$\alpha _{1}=\alpha _{4}=0$ and $\gamma _{2}=\gamma _{3}=0$, case 3. $\alpha
_{2}=\alpha _{3}=0$ and $\gamma _{1}=\gamma _{4}=0$, case 4. $\alpha
_{2}=\alpha _{3}=0$ and $\gamma _{1}=\gamma _{4}=0$. From the above four
cases, any state, which is connected with $L_{abc_{2}}$ with $abc\neq 0$, $%
a=b$ and $a\neq \pm c$ by SLOCC, satisfies the above property 3.4:

\subsubsection{Subsubfamily $L_{abc_{2}}$\ with $abc\neq 0$ and $a=-b$}

When $a\neq \pm c$, $L_{abc_{2}}$ becomes $L_{abc_{2}}(a=-b)=a$ $%
(|0011\rangle +|1100\rangle )+c$\ $($\ $|0101\rangle +|1010\rangle
)+|0110\rangle $. For any state which is connected with $L_{abc_{2}}$ with $%
abc\neq 0$, $a=-b$ and $a\neq \pm c$ by SLOCC, $F_{i}$ ($i=1$, ..., $8$)
satisfy Eq. (\ref{Labc2-F-2}) and the following $D_{1}$, $D_{2}$, and $D_{3}$
hold.

$D_{1}=-a(a^{2}-c^{2})\alpha _{1}\alpha _{3}\gamma _{2}\gamma _{4}\det
(\alpha )\det^{2}(\beta )\det (\gamma )\det^{2}(\delta )$,

$D_{2}=-c(a^{2}-c^{2})\alpha _{1}\alpha _{3}\beta _{2}\beta _{4}\det (\alpha
)\det (\beta )\det^{2}(\gamma )\det^{2}(\delta )$,

$D_{3}=ac\det (\alpha )\det^{2}(\beta )\det^{2}(\gamma )\det (\delta )\ast $

$(a^{2}\alpha _{2}\alpha _{4}\delta _{1}\delta _{3}+c^{2}\alpha _{2}\alpha
_{4}\delta _{1}\delta _{3}+ac\alpha _{2}\alpha _{3}\delta _{2}\delta
_{3}+ac\alpha _{1}\alpha _{4}\delta _{2}\delta _{3}$

$+ac\alpha _{2}\alpha _{3}\delta _{1}\delta _{4}+ac\alpha _{1}\alpha
_{4}\delta _{1}\delta _{4}+a^{2}\alpha _{1}\alpha _{3}\delta _{2}\delta
_{4}+c^{2}\alpha _{1}\alpha _{3}\delta _{2}\delta _{4})$.

Property 3.5 also holds. We can derive the following properties 4.1-4.4 for
each state which is connected with $L_{abc_{2}}$ with $abc\neq 0$, $a=-b$
and $a\neq \pm c$\ by SLOCC.

Property 4.1.\textbf{\ }If $F_{3}=F_{4}=0$, then $D_{2}=0$.

Property 4.2\textbf{. }If $F_{1}=F_{2}=0$, then $D_{1}=D_{2}=0$.

Property 4.3.\textbf{\ }If $F_{5}=F_{6}=0$, then $D_{1}=0.$

Property 4.4.\textbf{\ }If $F_{1}=F_{2}=F_{7}=F_{8}=0$, then $D_{3}\neq 0$.

From Eq. (\ref{Labc2-state-F}), for the state $L_{abc_{2}}(a=-b\neq 0)$, $%
F_{i}=0$, $i=1,2,3,4,5,6,7,8$, and $D_{1}=D_{2}=0$. Therefore, the state $%
L_{abc_{2}}(a=-b\neq 0)$ does not satisfy Property 3.4. It means that
classes B3.2 and B3.3 are different from classes B2.2 and B2.3, respectively.

\subsubsection{Subsubfamily $L_{abc_{2}}$\ with $abc\neq 0$ and $a\neq \pm b$%
, but $c=\pm a$ or $c=\pm b$}

By discussion in Sec. 4.2.3, we only need to consider $c=a$ here.

When $c=a$, $L_{abc_{2}}$ becomes $\frac{a+b}{2}(|0000\rangle +|1111\rangle
)+\frac{a-b}{2}(|0011\rangle +|1100\rangle )+a(\ |0101\rangle +|1010\rangle
)+|0110\rangle $. Thus, each state in this subsubfamily satisfies the
following equations.

$F_{1}=a(a^{2}-b^{2})\alpha _{1}^{4}P$, $F_{2}=a(a^{2}-b^{2})\allowbreak
\alpha _{3}^{4}P$, $F_{3}=a(a^{2}-b^{2})\beta _{2}^{4}Q$, $%
F_{4}=a(a^{2}-b^{2})\beta _{4}^{4}Q$, $F_{5}=a(a^{2}-b^{2})\gamma _{2}^{4}R$%
, $F_{6}=a(a^{2}-b^{2})\gamma _{4}^{4}R$, $F_{7}=a(a^{2}-b^{2})\delta
_{1}^{4}S$, $F_{8}=a(a^{2}-b^{2})\allowbreak \delta _{3}^{4}S$.

We omit the complicated expressions of $F_{9}$, $F_{10}$, $D_{1}$, $D_{2}$,
and $D_{3}$.

It is plain to see that $F_{i}$ satisfy the following inequality.

\begin{equation}
\left\vert F_{i}\right\vert +\left\vert F_{i+1}\right\vert \neq 0,i=1,3,5,7%
\text{.}  \label{inequality-1}
\end{equation}

We also have the following properties for each state which is connected with 
$L_{abc_{2}}$ with $abc\neq 0$, $a\neq \pm b$, and $c=a$ by SLOCC.

Property 5.1.\textbf{\ }If $F_{1}F_{2}=0$ and $F_{5}F_{6}=0$, then $%
D_{1}\neq 0$.

Property 5.2.\textbf{\ }If $F_{1}F_{2}=0$ and $F_{3}F_{4}=0$, then $%
D_{2}\neq 0$.

Property 5.3.\textbf{\ }If $F_{1}F_{2}=0$ and $F_{7}F_{8}=0$, then $%
D_{3}\neq 0$.

Property 5.4.\textbf{\ }If $F_{2}=F_{3}=0$, then $F_{9}\neq 0$ and $%
F_{10}\neq 0$.

In Eq. (\ref{Labc2-state-F}), $F_{i}$ and $D_{2}$ obtained by letting $a=\pm
b$ do not satisfy Properties 5.2. It implies that classes B2.2, B2.3, B3.2
and B3.3 are different from classes B4.1 and B4.2.

\subsubsection{Subsubfamily $L_{abc_{2}}$ with $abc\neq 0$, and $x\neq \pm y$%
, where $x$, $y\in \{a$, $b$, $c\}$}

Each state in this subsubfamily satisfies the following.

$F_{1}=\alpha _{1}^{2}[c(a^{2}-b^{2})\alpha
_{1}^{2}+(a^{2}-c^{2})(b^{2}-c^{2})\alpha _{2}^{2}]P$,

$F_{2}=\alpha _{3}^{2}[c(a^{2}-b^{2})\alpha
_{3}^{2}+(a^{2}-c^{2})(b^{2}-c^{2})\alpha _{4}^{2}]P$,

$F_{3}=\beta _{2}^{2}[c(a^{2}-b^{2})\beta
_{2}^{2}+(a^{2}-c^{2})(b^{2}-c^{2})\beta _{1}^{2}]P$,

$F_{4}=\beta _{4}^{2}[c(a^{2}-b^{2})\beta
_{4}^{2}+(a^{2}-c^{2})(b^{2}-c^{2})\beta _{3}^{2}]P$,

$F_{5}=\gamma _{2}^{2}[c(a^{2}-b^{2})\gamma
_{2}^{2}+(a^{2}-c^{2})(b^{2}-c^{2})\gamma _{1}^{2}]P$,

$F_{6}=\gamma _{4}^{2}[c(a^{2}-b^{2})\gamma
_{4}^{2}+(a^{2}-c^{2})(b^{2}-c^{2})\gamma _{3}^{2}]P$,

$F_{7}=\delta _{1}^{2}[c(a^{2}-b^{2})\delta
_{1}^{2}+(a^{2}-c^{2})(b^{2}-c^{2})\delta _{2}^{2}]P$,

$F_{8}=\delta _{3}^{2}[c(a^{2}-b^{2})\delta
_{3}^{2}+(a^{2}-c^{2})(b^{2}-c^{2})\delta _{4}^{2}]P$,

$\mathcal{I}=\frac{1}{2}(a^{2}+b^{2}+2c^{2})T$.

We want to show that some states in this subsubfamily violates Eq. (\ref%
{inequality-1}). So, this subsubfamily is inequivalent to the subsubfamily $%
L_{abc_{2}}$\ with $abc\neq 0$ and $a\neq \pm b$, but $c=\pm a$ or $c=\pm b$
under SLOCC. The following is our argument.

For the operator $\alpha $, let $\alpha _{1}=0$, $\alpha _{2}\neq 0$, $%
\alpha _{3}^{2}=\frac{(a^{2}-c^{2})(c^{2}-b^{2})}{c(a^{2}-b^{2})}\alpha
_{4}^{2}$, and $\alpha _{4}\neq 0$. Clearly, $\det (\alpha )\neq 0$, but $%
F_{1}=F_{2}=0$. So, it violates Eq. (\ref{inequality-1}).

\subsection{Subfamily $L_{abc_{2}}$\ with $c\neq 0$ and $ab=0$}

There are three subsubfamilies. They are the subsubfamily $L_{abc_{2}}$\
with $c\neq 0$ and $a=b=0$, the subsubfamily $L_{abc_{2}}$\ with $c\neq 0$
and $a=0$ and $b=c$, and the subsubfamily $L_{abc_{2}}$\ with $bc\neq 0$ and 
$a=0$ and $b\neq \pm c$.

\subsubsection{Subsubfamily $L_{abc_{2}}$\ with $c\neq 0$ and $a=b=0$}

$L_{abc_{2}}$ becomes $L_{abc_{2}}(a=b=0)=$\ $c$\ $($\ $|0101\rangle
+|1010\rangle )+|0110\rangle $. Each state, which is connected $%
L_{abc_{2}}(a=b=0)$ by SLOCC, satisfies the following $D_{i}$ and $F_{i}$.

$D_{1}=D_{3}=0$, $D_{2}=c^{3}\alpha _{1}\alpha _{3}\beta _{2}\beta _{4}\det
(\alpha )\det (\beta )\det^{2}(\gamma )\det^{2}(\delta )$,

$F_{i}$ ($i=1$, ..., $8$) can be obtained from Eq. (\ref{Labc2-F-2}) by
letting $a=b=0$.

\ $F_{9}=c^{4}(\alpha _{1}\alpha _{4}\beta _{2}\beta _{3}-\alpha _{2}\alpha
_{3}\beta _{1}\beta _{4})^{2}\det^{2}(\gamma )\det^{2}(\delta )$,

$F_{10}=c^{4}(\alpha _{2}\alpha _{3}\beta _{2}\beta _{3}-\alpha _{1}\alpha
_{4}\beta _{1}\beta _{4})^{2}\det^{2}(\gamma )\det^{2}(\delta )$.

It is easy to know that the above $F_{i}$ satisfy Properties 2.1 and 2.2.

Each state, which is connected with $L_{abc_{2}}$\ with $c\neq 0$ and $a=b=0$
under SLOCC, satisfies the following Properties 6.1 and 6.2.

Property 6.1. If $F_{1}=F_{2}=0$ or $F_{3}=F_{4}=0$, then $D_{2}=0$.

Property 6.2. If $F_{1}F_{2}=0$ and $F_{3}F_{4}=0$, then ${\small F}_{9}%
{\small =0}\wedge {\small F}_{10}{\small \neq 0}$ or ${\small F}_{9}{\small %
\neq 0}\wedge {\small F}_{10}{\small =0}.$

Remark 1. Classes B1.5 and B5.1 are different though they have the same
information for $\mathcal{I}$,$\ D_{1}$, $D_{2}$, and $D_{3}$.

We argue remark 1 as follows. From Eq. (\ref{Labc2-Fi-1}), when $ab=0,$ any
state in class B1.5 satisfies $F_{i}=0$, $i=1,...,10$. But a representative
of class B5.1 is $c(\ |0101\rangle +|1010\rangle )+|0110\rangle $, for which 
$F_{10}=c^{4}\neq 0$ from Eq. (\ref{Labc2-state-F}). Therefore classes B1.5
and B5.1 are different.

\subsubsection{Subsubfamily $L_{abc_{2}}$\ with $c\neq 0$, $a=0$, and $b=c$}

This subsubfamily is a true SLOCC entanglement class.

For each state of this class, we have the following $F_{i}$.

$F_{1}=-b^{3}\alpha _{1}^{4}P$, $F_{2}=-b^{3}\allowbreak \alpha _{3}^{4}P$, $%
F_{3}=-b^{3}\beta _{2}^{4}Q$, $F_{4}=-b^{3}\beta _{4}^{4}Q$,

$F_{5}=-b^{3}\gamma _{2}^{4}R$, $F_{6}=-b^{3}\gamma _{4}^{4}R$, $%
F_{7}=-b^{3}\delta _{1}^{4}S$, $F_{8}=-b^{3}\allowbreak \delta _{3}^{4}S$.

$F_{i}$ satisfy Eq. (\ref{inequality-1}). Therefore, this class is different
from the classes obtained from $c=0$ and the classes obtained from $c\neq
0\wedge a=\pm b$ because the $F_{i}$ ($i=1$, ..., $8$) of the latter
representative states vanish. See Eq. (\ref{Labc2-state-F}). We also have
the following properties.

Properties 5.1, 5.2, and 5.3 hold for this case.

Each state, which is connected with $L_{abc_{2}}$\ with $c\neq 0$ and $a=0$
and $b=c$ under SLOCC, satisfies the following property 6.3

Property 6.3\textbf{. }If $F_{2}=F_{3}=0$, then $F_{9}=0$ and $F_{10}\neq 0$.

From Eq. (\ref{Labc2-state-F}), it is obvious that the representative state
with $a=0$ and $b=c\neq 0$ does not satisfy property 5.4. Hence, the class
B5.2 is different from classes B4.1 and B4.2, respectively.

For the state $L_{abc_{2}}$ with $c=0$ and $ab\neq 0$, we obtain $F_{9}\neq
0 $ and $F_{10}=0$ from Eq. (\ref{Labc2-state-F}). Thus, it violates
property 6.3. It means that classes B1.3 and B1.4 are different from the
class B5.2.

From Eq. (\ref{Labc2-state-F}), the representative state $L_{abc_{2}}$ with $%
abc\neq 0$ does not satisfy property 6.2 or 6.3. Hence, the classes in the
subfamily $L_{abc_{2}}$ with $abc\neq 0$ are different from the classes B5.1
and B5.2, respectively.

\subsubsection{Subsubfamily $L_{abc_{2}}$\ with $bc\neq 0$ and $a=0$ and $%
b\neq \pm c$}

Each state of this subsubfamily satisfies the following equations.

$\mathcal{I}=(b^{2}+2c^{2})/2\ast T$,

$F_{1}=-c\alpha _{1}^{2}(b^{2}\alpha _{1}^{2}+c(b^{2}-c^{2})\alpha
_{2}^{2})\ast P$, $F_{2}=-c\alpha _{3}^{2}(b^{2}\alpha
_{3}^{2}+c(b^{2}-c^{2})\alpha _{4}^{2})\ast P$,

$F_{3}=-c\beta _{2}^{2}(b^{2}\beta _{2}^{2}+c(b^{2}-c^{2})\beta
_{1}^{2})\ast Q$, $F_{4}=-c\beta _{4}^{2}(b^{2}\beta
_{4}^{2}+c(b^{2}-c^{2})\beta _{3}^{2})\ast Q$,

$F_{5}=-c\gamma _{2}^{2}(b^{2}\gamma _{2}^{2}+c(b^{2}-c^{2})\gamma
_{1}^{2})\ast R$, $F_{6}=-c\gamma _{4}^{2}(b^{2}\gamma
_{4}^{2}+c(b^{2}-c^{2})\gamma _{3}^{2})\ast R$,

$F_{7}=-c\delta _{1}^{2}(b^{2}\delta _{1}^{2}+c(b^{2}-c^{2})\delta
_{2}^{2})\ast S$, $F_{8}=-c\delta _{3}^{2}(b^{2}\delta
_{3}^{2}+c(b^{2}-c^{2})\delta _{4}^{2})\ast S$.

We can choose $\alpha _{1}=0$, $\alpha _{2}\neq 0$, $\alpha _{4}\neq 0$, and 
$\alpha _{3}^{2}=\frac{c(c^{2}-b^{2})}{b^{2}}\alpha _{4}^{2}$ such that $%
\det (\alpha )\neq 0$ but $F_{1}=F_{2}=0$. This violates Eq. (\ref%
{inequality-1}). Therefore this subsubfamily is different from the
subsubfamily $L_{abc_{2}}$\ with $c\neq 0$ and $a=0$ and $b=c$.

\ \ \ \ \


\begin{thebibliography}{99}
\bibitem{Dur} W. D$\ddot{u}$r, G.Vidal and J.I. Cirac (2000), Three qubits
can be entangled in two inequivalent ways, Phys. Rev. A. 62, 062314.

\bibitem{Moor2} F. Verstraete, J.Dehaene, B. De Moor and H. Verschelde
(2002), Four qubits can be entangled in nine different ways,\ Phys. Rev. A.
65, 052112.

\bibitem{Lamata1} L. Lamata, J. Le\'{o}n, D. Salgado and E. Solano (2006),
Inductive entanglement classification of four qubits under SLOCC, Phys. Rev.
A. 74 , 052336 .

\bibitem{Lamata2} L. Lamata, J. Le\'{o}n, D. Salgado and E. Solano (2007),
Inductive classification of multipartite entanglement under SLOCC,\ Phys.
Rev. A. 75, 022318. Also, see quant-ph/0610233.

\bibitem{LDF} D. Li, X. Li, H. Huang and X. Li (2008), Necessary and
sufficient conditions of separability

for multipartite pure states,\ Commun. Theor. Phys., Vol. 49, 1211-1216.
Submitted to Phys. Rev. Lett.\ (Sep. 2004), the paper No. LV9637. Late,
quant-ph/0604147.\ 

\bibitem{LDF07a} D. Li, X. Li, H. Huang and X. Li (2007), Classification of
four-qubit states by means of a stochastic operation and the classical
communication invariant and semi-invariants,\ Phys. Rev. A 76, 052311.\ Also
see,\ quant-ph/ 0701032.

\bibitem{Miyake03} A. Miyake (2003), Classification of multipartite
entangled states by multidimensional determinants,\ Phys. Rev. A. 67, 012108
.

\bibitem{LDF-PLA} D. Li, X. Li, H. Huang and X. Li (2006), Simple criteria
for the SLOCC classification,\ Phys. Lett. A 359, 428.

\bibitem{LDF07} D. Li, X. Li, H. Huang and X. Li (2007), Stochastic local
operations and classical communication invariant and the residual
entanglement for n qubits,\ Phys. Rev. A 76, 032304. Also see,
quant-ph/0704.2087.
\end{thebibliography}
\end{document}